\documentclass[structabstract]{aa}
\usepackage{txfonts}
\usepackage[utf8]{inputenc}
\usepackage[german,english]{babel}
\usepackage{natbib}
\usepackage{graphicx}
\usepackage{longtable}
\usepackage{lmodern}
\usepackage{amstext}
\usepackage{fp} 
\newcommand{\degree}{^\circ}
\newcommand{\activestars}{40.661 }
\newcommand{\numonefilter}{24.124 }
\newcommand{\numtwofilter}{18.619 }
\newcommand{\oneper}{5505 }
\newcommand{\MPone}{16.3 }
\newcommand{\MPtwo}{13.3 }
\newcommand{\sigPone}{10.1 }
\newcommand{\sigPtwo}{7.3 }

\newcommand{\dom}{\Delta\Omega}
\newcommand{\Msun}{M_{\odot}}
\newcommand{\one}{1}
\newcommand{\two}{2}
\newcommand{\three}{3}
\newcommand{\four}{4}
\newcommand{\five}{5}
\newcommand{\six}{6}
\newcommand{\seven}{7}
\newcommand{\eight}{8}

\FPeval{\actperc}{round(40661/165548*100,1)} 
\FPeval{\twoper}{round(\numtwofilter/\numonefilter*100,1)}

\begin{document}

\title{Rotation and differential rotation of active Kepler stars}

\author{ Timo Reinhold\inst{1} \and Ansgar Reiners\inst{1} \and Gibor Basri\inst{2} }

\offprints{T. Reinhold, \\ \email{reinhold@astro.physik.uni-goettingen.de} }

\institute{Institut für Astrophysik, Universität Göttingen, Friedrich-Hund-Platz 1, 37077 Göttingen, Germany\and Astronomy Department, University of California, Berkeley, CA 94720, USA}

\date{Received day month year / Accepted day month year}

\abstract
{
}
{
We present rotation periods for thousands of active stars in the Kepler field derived from Q3 data. In most cases a second period close to the rotation period was detected, which we interpreted as surface differential rotation (DR).
}
{
Active stars were selected from the whole sample using the range of the variability amplitude. To detect different periods in the light curves we used the Lomb-Scargle periodogram in a pre-whitening approach to achieve parameters for a global sine fit. The most dominant periods from the fit were ascribed to different surface rotation periods, but spot evolution could also play a role. Due to the large number of stars the period errors were estimated in a statistical way. We thus cannot exclude the existence of false positives among our periods.
}
{
In our sample of 40.661 active stars we found 24.124 rotation periods $P_1$ between 0.5 and 45 days. The distribution of stars with $0.5 < B-V < 1.0$ and ages derived from angular momentum evolution that are younger than 300\,Myr is consistent with a constant star-formation rate. A second period $P_2$ within $\pm\,30$\% of the rotation period $P_1$ was found in 18.619 stars (77.2\%). Attributing these two periods to DR we found that the relative shear $\alpha=\dom/\Omega$ increases with rotation period, and slightly decreases with effective temperature. The absolute shear $\dom$ slightly increases between $T_{\text{eff}}=3500-6000$\,K. Above 6000\,K $\dom$ shows much larger scatter. The dependence of $\dom$ on rotation period is weak over a large period range.
}
{
Latitudinal differential rotation measured for the first time in more than 18.000 stars provides a comprehensive picture of stellar surface shear, consistent with major predictions from mean-field theory. To what extent our observations are prone to false positives and selection bias is not fully explored, and needs to be addressed using more Kepler data.
}

\maketitle 

\section{Introduction}\label{intro}
The interplay of stellar rotation and convection is the origin of various stellar activity phenomena. For main sequence stars the rotation rate strongly depends on the stellar age. Due to rotational braking stars lose angular momentum over the time and slow down. \citet{Skumanich} empirically found the relation that the stars' rotational velocity is proportional to the inverse square root of its age: $v_{rot} \propto t^{-1/2}$. \citet{Barnes2003} shows that this relation holds for open cluster and Mount Wilson stars, furthermore providing a color dependence of the rotation period. \citet{Irwin2011} measure rotation periods for stars with masses below 0.35\,$\Msun$, finding some exceptionally fast and slow rotators. These stars do not follow the color-period relation from \citet{Barnes2003} but they can be explained by a radius-dependent braking efficiency \citep{Reiners2012}. Nowadays, a method called \textit{gyrochronology} \citep{Barnes2007} is being developed using the above Skumanich's relation in the opposite way to infer stellar ages from the rotation rate. Moreover the rotation rate strongly correlates with CaII emission it can be used as a measure of stellar activity. A relation between these properties is often called \textit{age-rotation-activity relation} \citep{Covey2010}. \\
The solar rotation profile is by no means uniform. Helioseismology reveals that the outer convective region shows large spread of rotation rates at different latitudes whereas the interior exhibits an almost constant rotation rate. Stars other than the Sun do not rotate rigidly as well. In early F-type stars a convection zone starts to form growing towards later spectral types. The Coriolis force acts on turbulence in the convection zone. Its back reaction redistributes angular momentum and changes the global rotation behavior, leading to differential rotation (hereafter DR) of the surface. A detailed theoretical description can be found in \citet{Kitchatinov2005}. DR of stellar interiors is studied by asteroseismology \citep{asteroseismology} but we are mainly interested in surface DR. On the Sun the equatorial region rotates faster than the poles, i.e. the angular velocity $\Omega$ depends on the latitude $\theta$. This latitudinal DR is usually described by the equation
\begin{equation}\label{law}
 \Omega(\theta)=\Omega_{\text{eq}}(1-\alpha_\odot \sin^2\theta)
\end{equation}
with $\Omega_{\text{eq}}$ being the angular velocity at the equator, and $\alpha_\odot=0.2$ the relative horizontal shear. In general $\alpha>0$ is known as solar-like DR, $\alpha<0$ is called anti solar-like DR, and $\alpha=0$ supplies rigid body rotation. 
The absolute shear $\dom$ between the equator and the pole is linked to $\alpha$ by the relation
\begin{equation}
 \dom=\Omega_{\text{eq}}-\Omega_{\text{pole}}=\alpha\,\Omega_{\text{eq}}.
\end{equation}
DR is believed to be one major ingredient of the driving mechanism of magnetic field generation on the Sun. Turbulent dynamos operating in other stars produce strong magnetic fields and are able to transform poloidal into toroidal fields, and vice versa. \citet{Morin2008} showed that the M4 dwarf V374 Peg exhibits a strong magnetic field showing only weak signatures of DR. This effect becomes even more important when stars become fully convective \citep{Morin2010}. Furthermore, the strength of DR varies with spectral type. \citet{Barnes2005} found that $\dom$
strongly increases with effective temperature. For temperatures above 6000\,K this trend was confirmed by other authors \citep{Reiners2006,Cameron2007}. This could be a hint towards different dynamo mechanisms, but the final role of DR is still not understood. \\
There are several ways to measure stellar rotation rates. The most common techniques are the long-term monitoring of photometric light curves yielding rotation periods from star spots, and the fit to spectral line profiles to measure rotational broadening ($v\sin i$). Other methods include line core variations in the CaII H \& K lines \citep{Baliunas1983,Gilliland1985}, and in eclipsing binaries the rotation rate can be measured by the Rossiter–McLaughlin effect or by ellipsoidal light variations. The rotation rate is a well-known quantity for tens of thousands of stars.  \\
DR is much harder to measure since surface features can only be resolved on the Sun. Nonetheless, star spots located at different latitudes are useful tracers for DR. Doppler Imaging tracks active regions and follows their migration over time to draw conclusions about the stellar rotation law. This method has successfully been used, see e.g. \citet{Donati1997,Cameron2002}. A different technique to measure DR is the Fourier transform method \citep{Reiners2002} analyzing the shapes of Doppler broadened line profiles. Following another approach, \citet{Lanza1993} simulated light curves of spotted stars and detected different periods by taking the Fourier-transform. \citet{Lucianne2012} fit an analytical spot model to synthetic light curves of spotted stars to see whether the model can break degeneracies in the light curve, especially accounting for the ability of determining the correct rotation periods, both in the presence and absence of DR. Analytical spot model were fit to real data, see e.g. \citet{MOST2006, Froehlich2009}, accounting for DR in the parameter space. Recently, this method was used for single Kepler light curves \citep{Frasca2011, Froehlich2012} where DR is the favorite explanation for the light curve shape. Asteroseismology provides another approach, explaining frequency splitting of global oscillations in terms of different latitudinal rotation rates \citep{Gizon2004}. \\
With the advent of the space missions CoRoT and Kepler photometric data of a vast number of stars was collected, continuously, simultaneously, and with an unprecedented precision. This enables us to study stellar variability of a large number of stars with variations of only millimagnitudes. Due to the plethora of data an automated classification for the different kinds of stellar variability is needed. Attempts were made to group the whole Kepler sample into known classes of variability like defined pulsation classes (e.g. RR Lyrae, $\delta$\,Scuti, etc.), rotation induced variability, binarity, and other groups \citep{Debosscher2011,Uytterhoeven2011}. In many cases, however, a unique classification was not possible. \\
The main goal of this paper is to provide rotation periods for a large fraction of the Kepler sample. Our special focus lies on the detection of a second period close to the rotation period as hint for DR. We apply the method from \citet{Reinhold2013} (hereafter Paper I) to derive different periods. Our analysis method is based on the  Lomb-Scargle periodogram \citep{Zechmeister2009} which has been successfully used to measure rotation periods for several CoRoT stars \citep{Affer2012}. \citet{Amy2013} have measured rotation periods utilizing an auto-correlation technique for the M dwarfs in the Kepler field, finding evidence for two different stellar populations due to a bimodal period distribution. \\
The paper is organized as follows. In Sec. \ref{reduction} the Kepler data and its reduction is discussed. Sec. \ref{sample} shows how the
active stars are selected from the whole sample. Sec. \ref{methods} summarizes the method developed in Paper I, with special focus on the period selection process. The rotation periods are presented in sec. \ref{rotation}, with a focus on DR in sec. \ref{DR} and \ref{DR1}. In sec. \ref{discussion} we compare our results to other observations and theoretical predictions. The last section contains the summary. DR beyond the imposed limits is shown in the appendix.

\section{Kepler data \& Reduction}\label{reduction}
Our analysis is based on the publicly available\footnote{http://archive.stsci.edu/pub/kepler/lightcurves/tarfiles/} Quarter 3 (Q3) long cadence data. Although plenty of data are available we restrict our analysis to one quarter because it is challenging to stitch different quarters together. Q3 has been chosen because it has less instrumental effects than earlier quarters, and carries a large number of targets (165.548 light curves in total). Generally each quarter is suitable for our purposes and we plan to use data from other quarters to validate our results and to see how periods change with time. \\
All Kepler light curves suffer from systematics hitting on various time scales. The most dominant one is due to differential velocity aberration manifesting in up- and downward low frequency trends of the light curves. Their removal is non-trivial \citep{McQuillan2011,Kinemuchi2012,Petigura2012} since one has to decide which trends are purely instrumental and which ones are due to true stellar variability. The uncorrected data are marked as SAP\_FLUX in the FITS files. The first pipeline available that tried to correct for instrumentals is called Presearch Data Conditioning (PDC) with the aim of finding planetary signals in the light curve. It was not very careful when removing variability from the light curves. Hence, in many cases true stellar variability has been removed. The next reduction pipeline PDC-MAP\footnote{MAP stands for Maximum A Posteriori which means that the parameters are estimated in a Bayesian way.} \citep{PDC-I,PDC-II} removes the so-called \textit{co-trending basis vectors} from the data. This pipeline removes the most common trends but keeps the stellar variability. All our calculations are based on this PDC-MAP data (SOC 8.3). The next section describes how active stars are selected from the whole Kepler sample.

\section{Sample selection}\label{sample}
\begin{figure}
  \resizebox{\hsize}{!}{\includegraphics{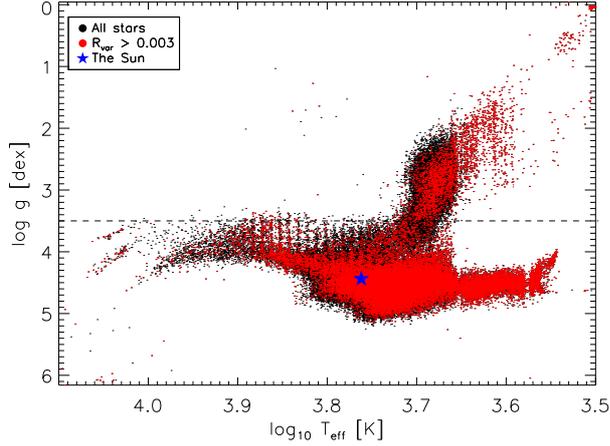}}
  \caption{Temperature vs. gravity of all Kepler Q3 stars (black) using KIC parameters. The active sample ($R_{var}\ge0.003$) is shown in red. The blue star marks the Sun which is shown for comparison. The dashed line marks $\log g = 3.5$ which was set to exclude giants in the following.}
  \label{HRD}
\end{figure}
\begin{figure}
  \resizebox{\hsize}{!}{\includegraphics{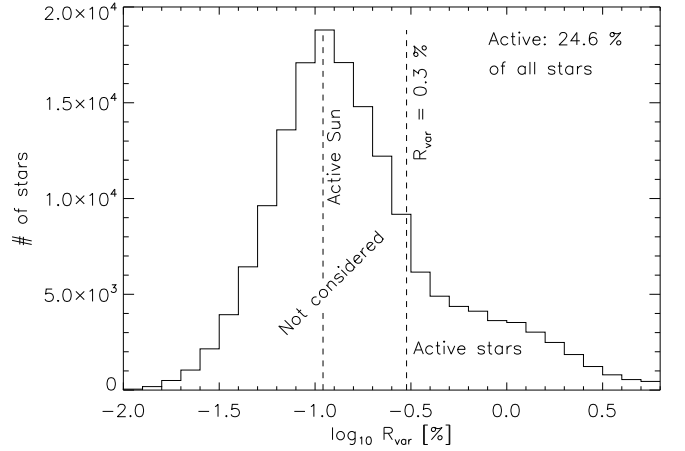}}
  \caption{Distribution of $R_{var}$ for all Kepler Q3 stars. \activestars active stars lie above the imposed limit $R_{var}\ge0.3\%$, which excludes more than 75\% of all stars.}
  \label{range}
\end{figure}

In Fig. \ref{HRD} we plot effective temperature vs. gravity of the whole Kepler Q3 sample (black dots), with the active stars shown in red. The selection of active stars is done automatically, i.e. without visual inspection of the light curves since the Kepler sample is huge. The active stars are selected using the so-called \textit{variability range} $R_{var}$ \citep{Basri2010,Basri2011}. The value is computed as follows: We sort the 4 hours boxcar smoothed differential light curve by amplitude, cut the upper and lower 5\%, and take the difference between top and bottom amplitude. This measure accounts for the intrinsic variability of the star, i.e. a variable star has a larger variability range than a quiet star. After visual inspection of several light curves we found that a suitable criterion is $R_{var}\ge0.003$ (3 parts per thousand). Most of the active stars populate the dwarf regime with $\log g \gtrsim 4$. The upper right corner shows a significant fraction of active cool stars with $\log g \lesssim 3$. Visual inspection of these low gravity stars reveals two groups of variability. The first one has very high ranges up to several percent, regular spot-like variations, and long periods. This might indicate spots or pulsations on giants, which we do not consider in this work. The second group exhibits irregular variability on short time scales which could be due to multiple mode pulsations. The Sun (blue star) is shown for comparison. All parameters have been taken from the Kepler Input Catalog (KIC). We see that the Kepler sample is strongly biased towards solar-like stars, but also a large giant branch ($\log \; g \lesssim 3.5$) is clearly visible. \\
We compared this value to total solar irradiance (TSI) data from the VIRGO instrument at the SOHO satellite. Using data from 1 Dec 1995 to 1 Sep 2011 we found that the maximum range was $max(R_{var,\odot})=0.0023$, with a mean of $\left\langle R_{var,\odot} \right\rangle=0.0011$ during solar maximum (Feb 20 1999 - Oct 21 2004). This value lies below our limit, thus all stars considered are more active than the active Sun. The variability range is our key measure to distinguish between active and quiet stars, i.e. all stars with $R_{var}$ above the upper limit will be called ``active'' although there is a large spread in their ranges. \activestars stars of the whole Kepler sample survive this criterion. The distribution of ranges is shown in Fig. \ref{range}. Only \actperc\% of all stars are considered as active, i.e. to these stars we will apply the analysis procedure from Paper I. \\

\section{Period determination}\label{methods} 
Stellar activity covers a wide field of different phenomena. Our main focus lies on the detection of periodic variability induced by dark spots co-rotating with the star. Since the spot periods constrain the stellar rotation law the detection of more than one dominant period is considered as hint for DR although there might be other effects able to mimic DR. Hence, the method from Paper I is applied to all active stars defined in the previous section. In sec. \ref{pw} we briefly recall our method to detect different periods in a light curve (for details we refer the reader to Paper I), and describe the selection process to achieve the physically meaningful periods (sec. \ref{selection}). In sec. \ref{zerocrossings} \& \ref{detlimsec} we apply different filters to the set of returned periods to make sure that rotation is their favorite explanation. Finally, we show three examples where DR is the favorite explanation for the light curve pattern (sec. \ref{examples}).

\subsection{Lomb-Scargle periodogram  \& Pre-whitening}\label{pw}
To detect periods in a light curve we used the generalized Lomb-Scargle periodogram \citep{Zechmeister2009}. To save computation time each light curve was binned to two hour bins. The lowest frequency is given by the inverse product of the time span ($\approx 90$ days) times an oversampling factor of 20, i.e. $f_{low}\approx1/(90\cdot20$\,d), which accounts for proper frequency sampling. Using a denser frequency sampling (factor of 30) did not change the results. The highest frequency is given by the Nyquist frequency using the above binning. The binning does not affect the period determination since we only considered periods longer than half a day (sec. \ref{selection}). \\
Computing the Lomb-Scargle periodogram is equivalent to fitting a sine wave to the data. Subtracting the sine function from the data and computing the periodogram of the residuals yields a second period. This \textit{pre-whitening} procedure was repeated five times to detect further periods in the light curve. Afterwards, all sine parameters were used as input for a global sine fit which is a simultaneous fit of the sum of all five sine waves allowing for all periods to vary. Using more pre-whitening steps results in a better fit but is computationally intensive and does not yield new significantly different periods. The enlarged set of returned periods makes it more difficult to assign a physical meaning to the individual periods (sec. \ref{selection}). Visual inspection of several light curves confirmed that the resulting fit was sufficient to detect significant rotation-induced periods. Three example light curves, the resulting fits, and the corresponding periodograms are discussed in sec. \ref{examples}.

\subsection{Period selection}\label{selection}
The next step was to select the most significant periods from the global sine fit, and to assign a physical meaning to them. We were interested in rotation periods of the star. The whole concept of only one rotation period is not quite exact, because one can only detect periodic variations caused by active regions located at certain latitudes. We think of either single spots or spot groups rotating over the visible hemisphere. If these regions are not located at the equator, then the equatorial rotation period remains unknown. Another problem arises from stars with a high spot coverage due to many small spots. If their surface distribution is inhomogeneous, then their light curves cannot be distinguished from a star with few active regions. Nevertheless, we used the first sine period from the global sine fit as the most significant period in the light curve. This period belongs to the highest power found in the pre-whitening process, and is therefore defined as one rotation period. In some cases the spots are located on opposite sides of the star, and the half period is what was detected. To minimize these alias periods we compared the first two periods from the global sine fit. If the difference of twice the first period and the second period is less than 5\%, then the algorithm selected the longer one which is likely the correct period. The period finally selected is our primary period $P_1$. \\
If the star rotates differentially, active regions have different velocities which manifests in a superposition of different periods in the light curve. To search for a second period close to $P_1$, we looked for a period $P_2$ within 30\% of $P_1$ in the remaining four sine periods. To estimate the relative surface shear of the two active regions we defined
\begin{equation}\label{alpha}
 \alpha:=|P_2-P_1|/P_{max}
\end{equation}
with $P_{max}=\max\{P_1,P_2\}$. Since we cannot tell from the light curve whether the star rotates solar-like or anti solar-like, we always assumed solar-like DR. To get closest to the total equator-to-pole shear we normalized the period difference by $P_{max}$. Equivalent to $P_{max}$ we defined $P_{min}=\min\{P_1,P_2\}$ to get closest to the equatorial period. $\alpha$ should hold the relations
\begin{equation}\label{limits}
\alpha_{min} \leq \alpha \leq \alpha_{max}.
\end{equation}
The upper limit $\alpha_{max}=0.3$ was a reasonable choice, because the solar value is $\alpha_\odot=0.2$, and the total amount of DR is unknown for stars other than the Sun. In appendix \ref{beyond} we showed that the general results were not affected by using different $\alpha_{max}$ values. The lower limit $\alpha_{min}=0.01$ accounts for the fixed frequency resolution of the periodograms. If there were more than one sine period satisfying both criteria, then the one with the second highest power in the pre-whitening process was chosen. If $P_1$ was an alias period, then only the remaining three sine periods were considered to look for a second period. \\
In contrast to previous studies we found that the highest peak of the initial periodogram was a bad measure to filter out active stars. The periodogram often detected a period longer than 90 days. These long periods remain doubtful because one cannot distinguish between remnants from the PDC-MAP pipeline and real long-term variability. Their peak height in the periodogram was similar to more reliable shorter periods, i.e. the peak height was highly biased by the data reduction. Thus, the variability range was the only measure we used. \\
Now, we make further restrictions to the derived periods to assure that these are really due to rotation. Using the variability range, we selected \activestars active stars from the whole sample. As seen in Fig. \ref{HRD} there were several active giants. Since we are mainly interested in rotation periods of main-sequence stars, the surface gravity was restricted to $\log g > 3.5$ marked by the black dashed line in Fig. \ref{HRD}. Further limits were applied to the period $P_1$ setting 0.5\,d $\leq P_1 \leq$ 45\,d. The lower limit should exclude pulsations, which basically occur on short time scales. The upper limit is approximately equal to half the time span of Q3. Since Kepler suffers from instrumental effects visible on timescales of the quarter duration, periods longer than 45 days remain doubtful. We cannot distinguish between long-term variability, and trends caused by the instrument, because the periods were selected automatically by our algorithm. There exists a certain fraction of stars with long periods nonetheless. We cannot be sure that some of them are also due to instrumental effects, so they should be treated with some caution. Furthermore, all derived periods $P_1$ for relevant stars are compared to the orbital periods $P_{orb}$ from the lists of eclipsing binaries\footnote{http://archive.stsci.edu/kepler/eclipsing\_binaries.html} and false positives\footnote{http://archive.stsci.edu/kepler/false\_positives.html}. If the relation $|P_1-P_{orb}|/P_{orb}<0.05$ holds, these periods were discarded. This limit excludes orbital periods, but we might lost some tidally locked systems.

\subsection{Zero crossings}\label{zerocrossings}
\begin{figure}
  \resizebox{\hsize}{!}{\includegraphics{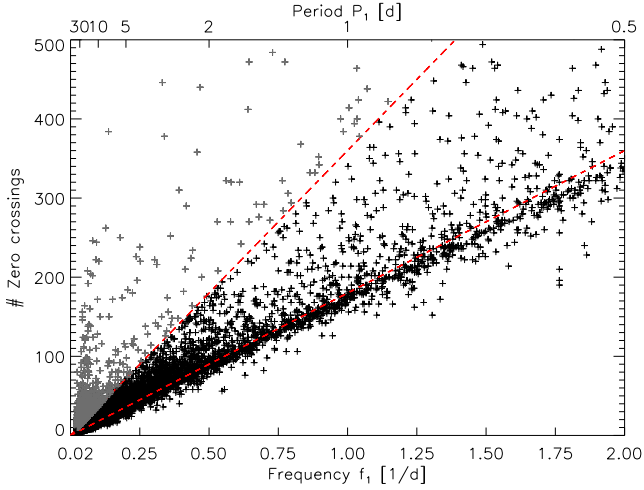}}
  \caption{Frequency $f_1$ vs. number of zero crossings in Q3. The lower red line equals the number of zero crossings expected for sine-like variations. Stars lying above the upper red line (gray symbols) were discarded, because they show more than twice zero crossings as expected.}
  \label{zeros}
\end{figure}

After setting several constraints to $P_1$, we applied a filter to achieve periods originating from rotating active regions by counting the number of zero crossings of each light curve. For many possible realizations of a spotted star, the light curve showed a sine-like variation with a defined number of zero crossings. Thus, a low number of crossing events was indicative for regular rotation-induced variations, whereas a high number of zero crossings was considered as a hint for stellar pulsations, and irregular variations. In this way we filtered out periods that did not originate from rotation. \\
A single sine wave has two zero crossings per period. Thus, the number of zero crossings in Q3 equals $N_{zero} = 90\cdot2/P_1=180\;f_1$ for a sine wave with a period $P_1$. To calculate $N_{zero}$ the light curves had to be smoothed. Since we were facing a variety of different rotation time scales (0.5--45 days), we could not apply the same smoothing width to all light curves. The very fast rotators can only be smoothed on a few hours to stay below the rotation period. For the very slow rotators other effects (like e.g. granulation) becomes dominant on these time scales so the smoothing time needs to be sufficiently longer. To account for this effect, we smoothed the light curves over timescales proportional to their period $P_1$. Since Kepler long cadence data consists of $\approx 30$ minutes integrations, we had $48 \cdot P_1$ data points in a certain period $P_1$. We smoothed the data using a boxcar average with a width of $4\cdot P_1$ data points, which turned out to be a suitable width for fast and slow rotators. We plot the frequency $f_1$ vs. the number of zero crossings in Q3 in Fig. \ref{zeros}.
The lower red line equals $N_{zero}=180\cdot f_1$ as predicted for sine like variations. The upper one equals $2N_{zero}=90\cdot4/P_1$. This upper limit accounts for the cases of two active longitudes located on opposite sides of the star, which produces up to four zero crossings per period. This trend was broken toward longer periods; the variations became irregular showing more zero crossings. A simple explanation is that the smoothing width was chosen too small. Another reason could be that star spots evolve on these long time scales ($P_1 > 20$ days) as seen on the Sun. Stars with too many zero crossings were considered as irregular, quiet, pulsators, or giants with poorly determined $\log g$. All stars lying above the upper red line (gray symbols) were discarded.

\subsection{Detection limit of $\alpha$}\label{detlimsec}
After filtering out periods by their number of crossing events, we focused on the cases where two periods were found. The reliability of a second period strongly depends on the separation of the periods $P_1$ and $P_2$ in the periodogram. If they lie very close it is hard to tell whether the small separation comes from a very small $\alpha$, or if it is due to artifacts from our method. After visual inspection of several light curves we found that the typical separation of the two periods should be larger than ten points on the frequency grid in the periodogram. Thus, we inverted the periods to achieve two frequencies $f_1=1/P_1$ and $f_2=1/P_2$. If their absolute separation was less than ten times the lowest frequency $f_{low}$ (sec. \ref{pw}), i.e.  $|f_1-f_2| < 10f_{low} \approx 0.0056$\,cycles\,d$^{-1}$, then the second period was discarded. \\
For each of the above limits imposed upon the set of periods, we might have lost real rotation periods. This was not preventable, because the analysis procedure was done automatically. All filters applied have decreased the number of false-positives significantly, yielding a condensed and reliable set of rotation periods. We found that \numonefilter stars survived all filters, i.e. they have a measured rotation period $P_1$. For \numtwofilter stars a second period $P_2$ was found. The following section shows example light curves, the fits achieved by our method, and the associated periodograms. It demonstrates the problem of determining a second period from the periodograms, and shows the importance of pre-whitening.

\subsection{Examples: Light curves, periodograms, and rotation periods}\label{examples}
\begin{figure*}
  \centering
  \includegraphics[width=17cm, bb=0 150 468 648]{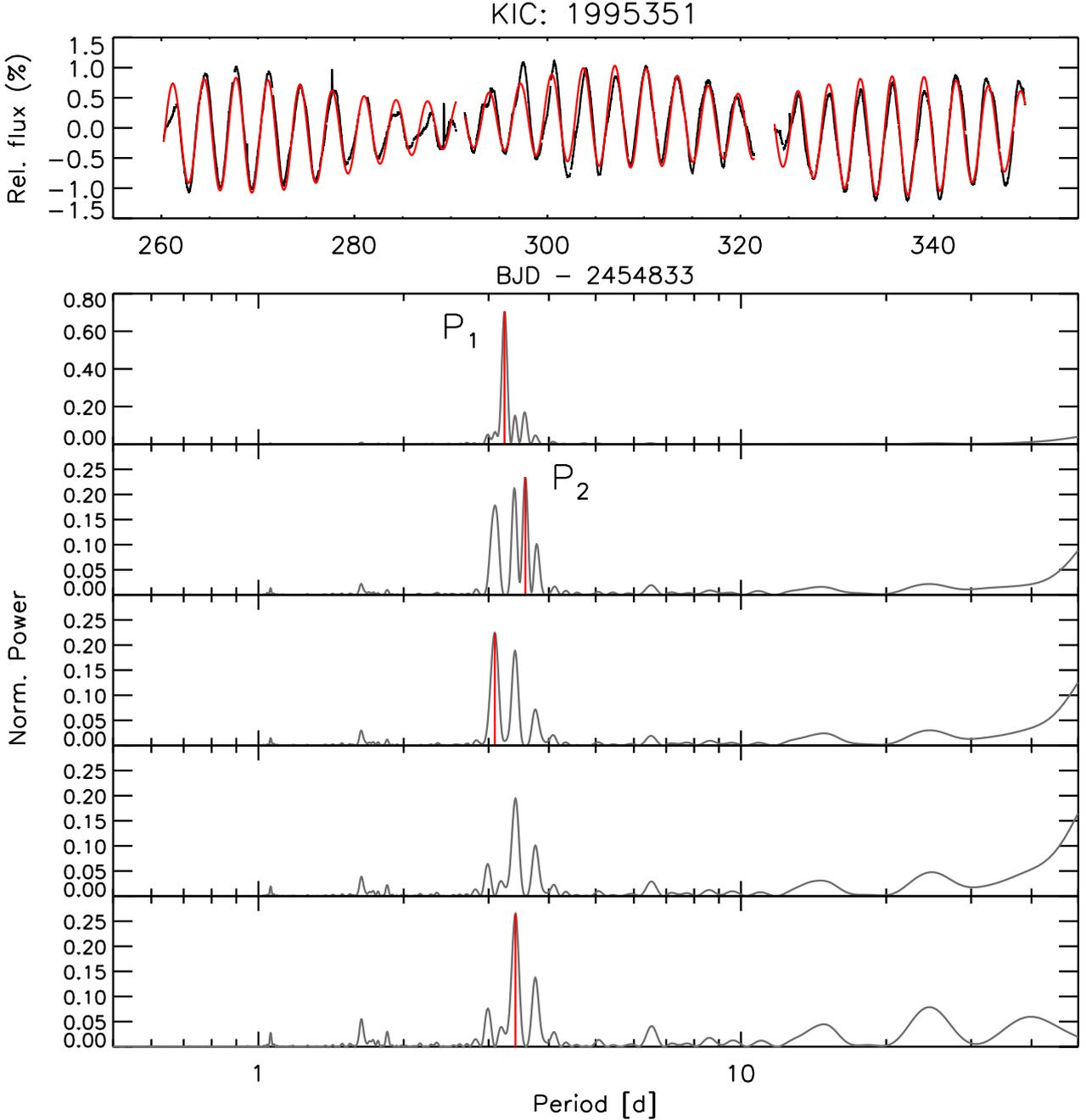}
  \caption{ \textit{Top Panel:} Light curve \& global fit of the star KIC 1995351. \textit{Lower Panel (top to bottom):} Periodograms 1--5. The vertical red lines indicate the highest peak in each periodogram. The periods $P_1=3.24$\,d and $P_2=3.57$\,d were selected by our method.}
  \label{kepler_lcs_perdgms}
\end{figure*}

We briefly address the problem of associating the returned periods from the pre-whitening to rotation periods of active regions on the star. Fig. \ref{kepler_lcs_perdgms} shows an example of a fast rotator, KIC 1995351, and the associated periodograms. The upper panel shows the light curve and the global sine fit of all five periods in red. The light curve shows a regular beating pattern most probably due to several active regions. The lower panel shows - from top to bottom - the first to the fifth periodogram. The highest peak in each periodogram was marked by a vertical red line. \\
The first periodogram reveals several distinct peaks close to the most significant one at 3.24 days, which was chosen as primary period $P_1$. In the second panel the periodogram of the residuals was taken. One clearly sees that the power of the peak around $P_1=3.24$ days basically dropped to zero, whereas the second and third highest peak from the initial periodogram now exhibit the highest powers. The period with the second highest power around $P_2=3.57$ days has been chosen. From the third to the fifth panel we see that there are more periods close to $P_1$ probably due to other active regions. In the fourth periodogram the vertical red line is not visible because the highest peak lies at 63.6 days, which is most probably an artifact from the data reduction. The periodogram fits long-term trends in the light curve with periods much longer the rotation periods. \\
\begin{figure*}
  \centering
  \includegraphics[width=17cm, bb=0 150 468 648]{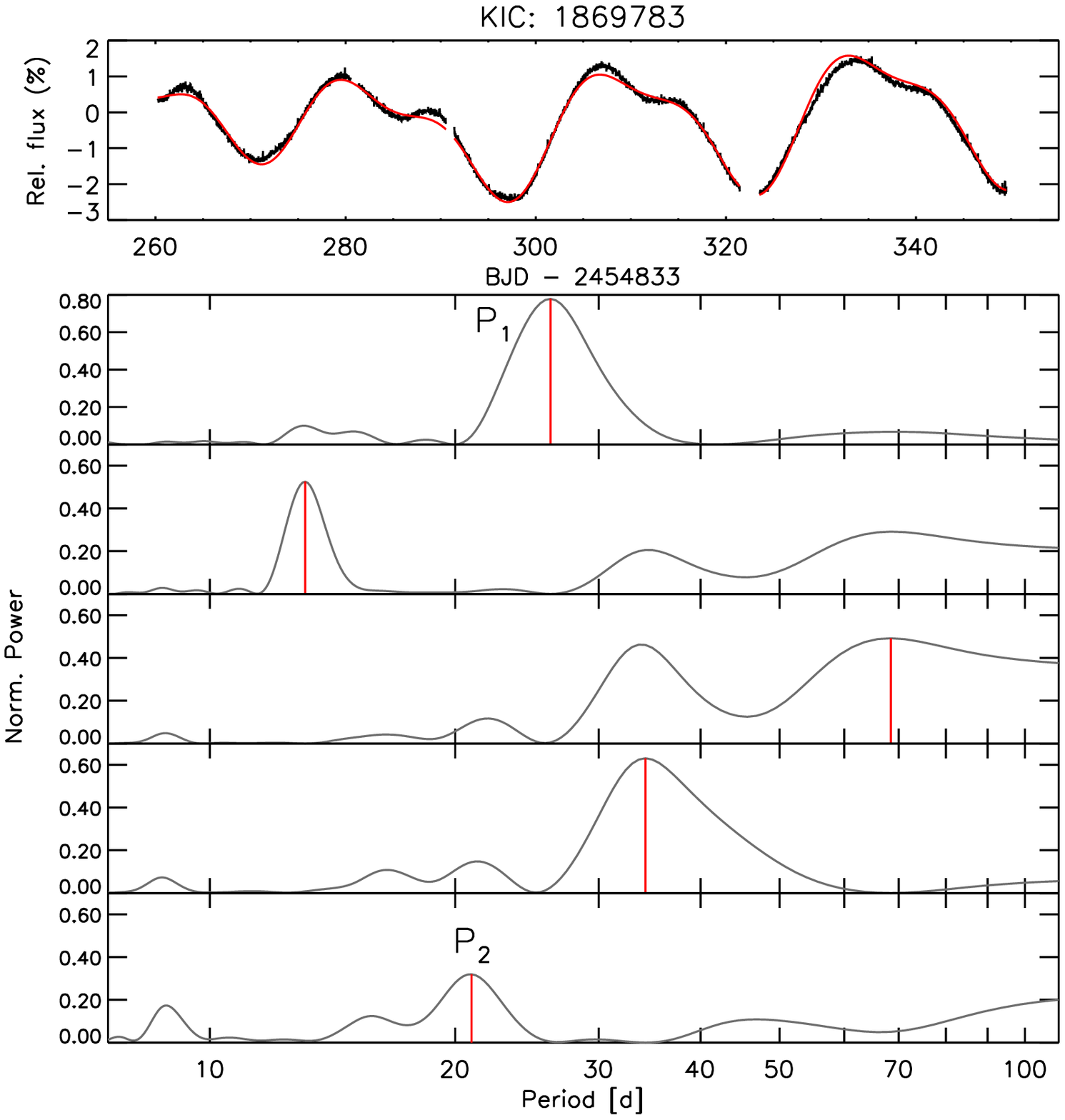}
  \caption{ \textit{Top Panel:} Light curve \& global fit of the star KIC 1869783. \textit{Lower Panel (top to bottom):} Periodograms 1--5. The vertical red lines indicate the highest peak in each periodogram. The periods $P_1=26.2$\,d and $P_2=20.9$\,d were selected by our method.}
  \label{kepler_lcs_perdgms1}
\end{figure*}
In Fig. \ref{kepler_lcs_perdgms1} we show the slow rotator KIC 1869783 and the associated periodograms. The panels are the same as in the previous plot. The light curve has a double-dip shape due to active regions located on opposite sides of the star. The initial periodogram shows a rather broad peak at 26.2 days which has been chosen as primary period $P_1$. The second periodogram shows a peak around 13.1 days which belongs the second active region on the opposite side of the star. From the light curve we can see how this region becomes shallower, and the primary region more pronounced. This traveling wave is usually interpreted as migrating star spots. The third periodogram has the highest peak around 68.5 days. Again, this peak most probably results from the data reduction. The fourth periodogram shows a strong peak around 34.2 days that has unfortunately been discarded by the selection process because it lies beyond 30\% of $P_1$. Only in the fifth pre-whitening step another period at 20.9 days has been found which was chosen as $P_2$ because it lies within 30\% of $P_1$. \\
These two examples demonstrate the process used to detect the most significant periods. The beating pattern seen in the light curve in Fig. \ref{kepler_lcs_perdgms} is no proof of DR, but that was the most probable explanation. The slow rotator shows active regions at different longitudes evolving with time. In this case it is not clear whether this was due to DR, or spot growth or decay. Both figures demonstrate that the fits to the light curves reproduce the main activity pattern. They could be improved by using more than five sine waves, but this would not change the two most significant periods. Furthermore, they point out the presence of a second period, and the difficulty to assign a physical meaning to them. \\
\begin{figure*}
  \centering
  \includegraphics[width=17cm, bb=0 150 468 648]{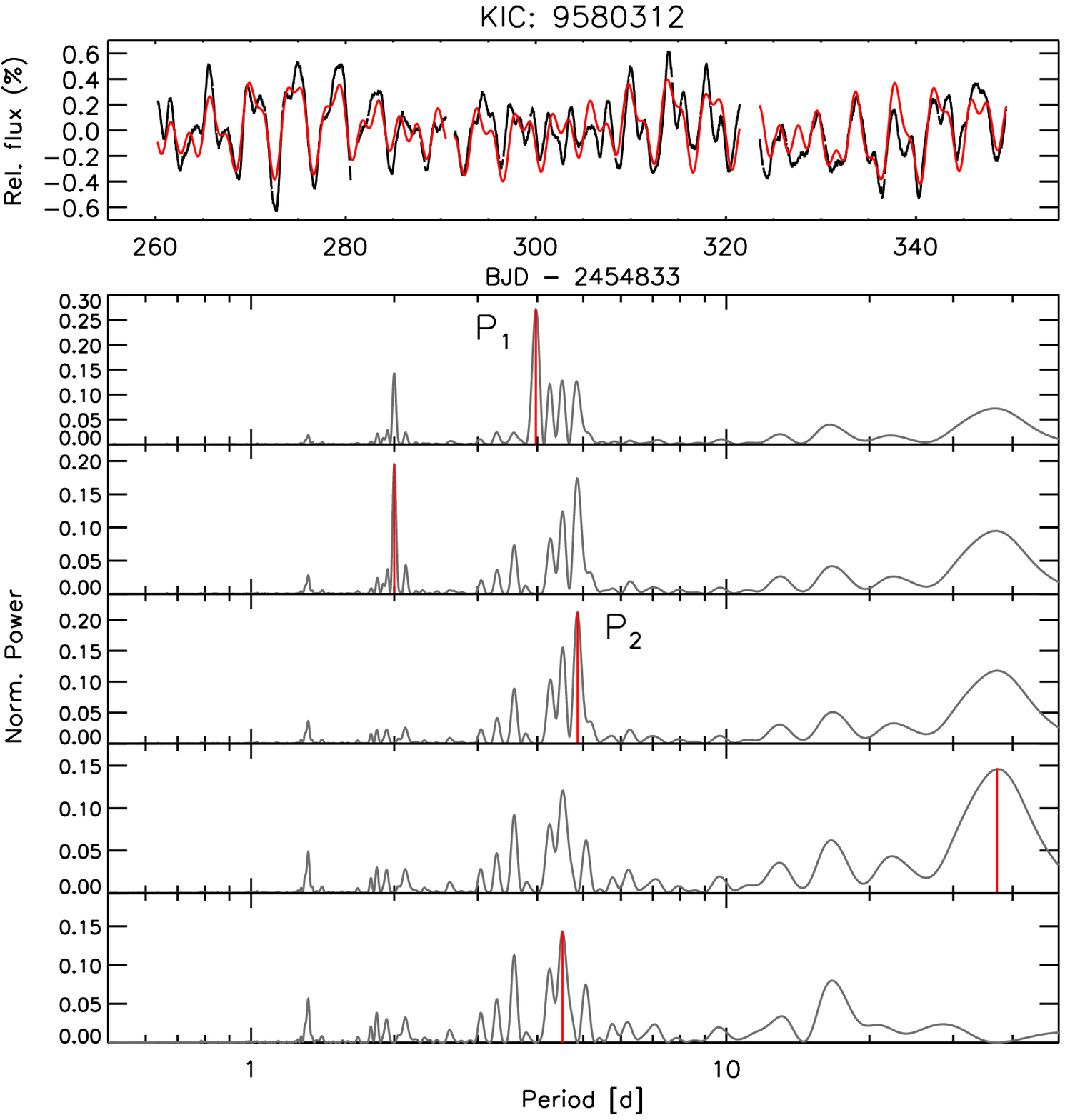}
  \caption{ \textit{Top Panel:} Light curve \& global fit of the star KIC 9580312. \textit{Lower Panel (top to bottom):} Periodograms 1--5. The vertical red lines indicate the highest peak in each periodogram. The periods $P_1=3.97$\,d and $P_2=4.86$\,d were selected by our method.}
  \label{Fstar}
\end{figure*}
Finally, we discuss the example of an F-type star in Fig. \ref{Fstar}. This particular star has an effective temperature of $T_{\text{eff}}=6504$\,K and a color index of $B-V=0.47$ (compare sec. braking), which roughly corresponds to a spectral type of F6V. F stars are known to exhibit very thin convection zones. Thus, it is questionable whether the variability in the light curve results from stars spots. Our pre-whitening analysis reveals the two most significant periods $P_1=3.97$\,d and $P_2=4.86$\,d. The second periodogram shows an alias of $P_1$ around 2 days. The highest peak in the fourth periodogram yields a period of 37.1 days that corresponds to the beating pattern. The fifth period found is located at 4.52 days which lies between $P_1$ and $P_2$. Fig. \ref{Fstar} clearly shows that the fit to the light curve is worse than in the other two examples, but still sufficient to reveal the two strongest periods. In this case, the beating pattern most probably results from differentially rotating spots, although this light curve shape was also observed in $\gamma$ Dor stars. We discuss this issue in more detail in sec. \ref{FP}.

\section{Results}\label{results}
In this section we present rotation periods of more than 20.000 Kepler stars. Sec. \ref{rotation} compares our results to previous measurements and sec. \ref{braking} shows that our periods are consistent with the concept of magnetic braking. Our main focus lies on the detection of DR, which is discussed in terms of relative and absolute shear in sec. \ref{DR} and \ref{DR1}, respectively. Finally, we estimate the number of false-positives, i.e. those periods mis-classified as rotation in sec. \ref{FP}, especially accounting for stars hotter than 6000\,K.

\subsection{Rotation Periods}\label{rotation}
\begin{figure}
  \resizebox{\hsize}{!}{\includegraphics{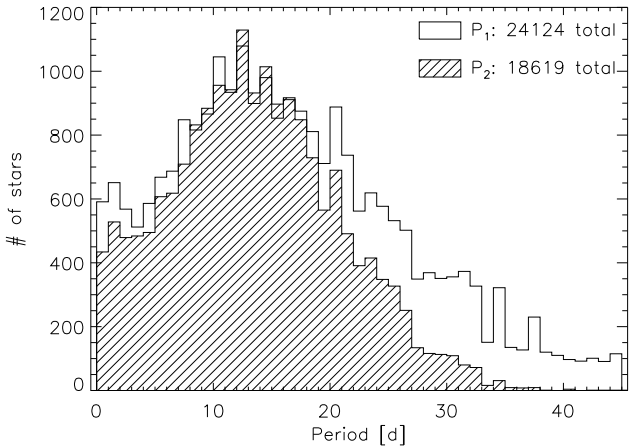}}
  \caption{Distribution of rotation periods for the \numonefilter stars with period $P_1$, and the \numtwofilter stars with second period $P_2$, with weighted means of \MPone and \MPtwo days, respectively. The distribution of $P_2$ falls off more rapidly towards 30 days, because instrumental effects currently hamper the detection of long periods.}
  \label{rotation_periods}
\end{figure}

In Fig. \ref{rotation_periods} we show the distribution of the rotation periods $P_1$ and $P_2$. We found $\left\langle P_1 \right\rangle=\MPone$\,d and $\left\langle P_2 \right\rangle=\MPtwo$\,d, with a spread of $\sigma(P_1)=\sigPone$\,d and $\sigma(P_2)=\sigPtwo$\,d, respectively. The distribution of $P_1$ slightly decreases toward longer periods, and levels off around $\approx 35$ days. Toward shorter periods a second peak between 0.5 and 2 days appears. The distribution of $P_2$ falls off more rapidly towards long periods. Most of the ``missing'' periods $P_2$ are greater than 20 days, so some of them might lie below our detection limit. Thus, the mean values of the distributions do not necessarily indicate that a second period is more likely be found for shorter periods. The dearth of slow rotators in both distributions is due to several reasons. The determination of long periods requires stable active regions on the star. On the Sun the spot lifetimes are of the order of the rotation period hampering the period detection. This does not need to be true for other stars though. Furthermore, long-term stellar variability and instrumental trends are currently difficult to distinguish in Kepler data. Both effects bias the distribution towards shorter periods. Our results apply primarily to periods shorter than about 30 days. Since we only analyzed one quarter so far, and applied an upper limit of 45 days, the distribution of slow rotators is not addressed in this study. 

\subsection{Rotational Braking}\label{braking}
\begin{figure*}
  \centering
  \includegraphics[width=17cm]{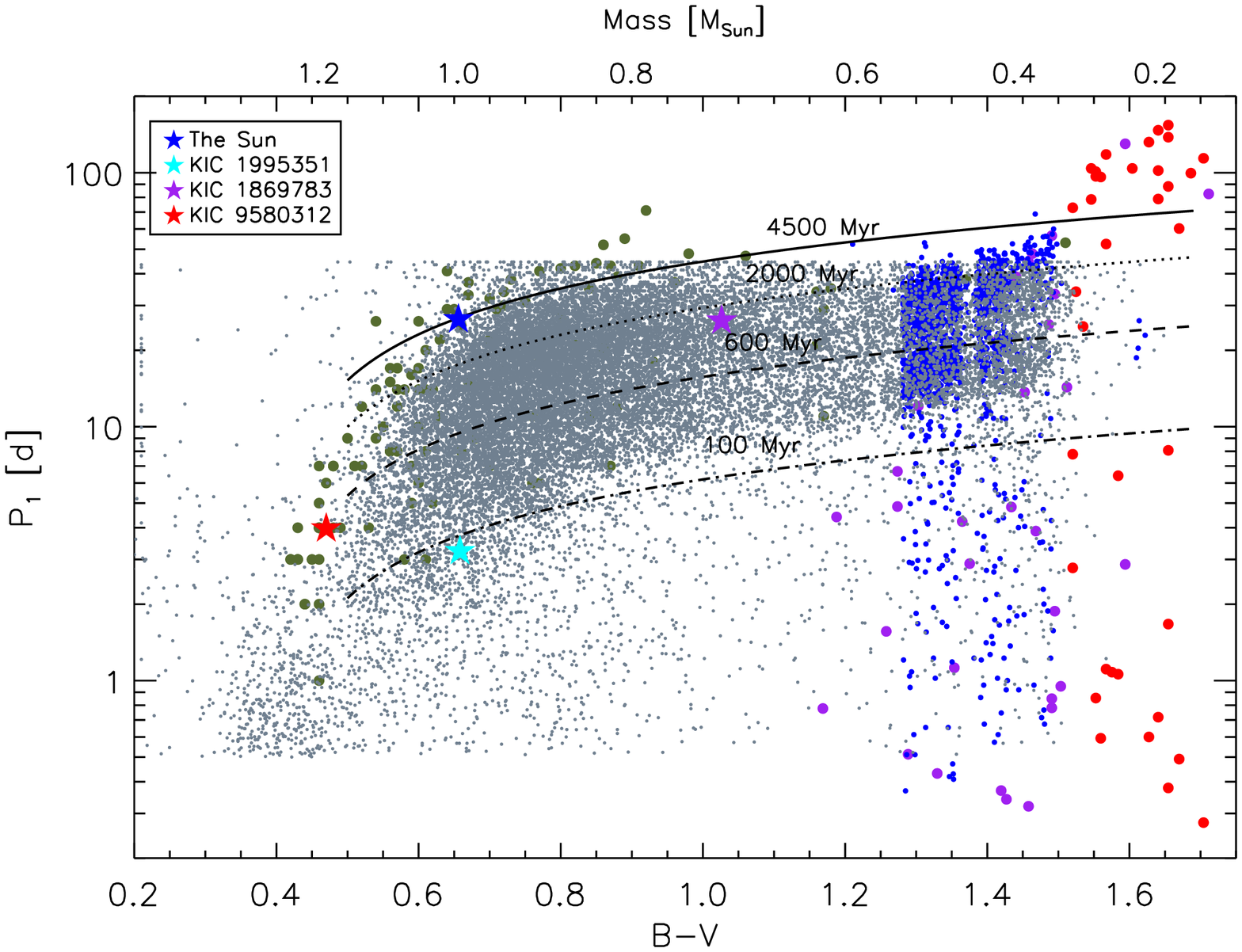}
  \caption{$B-V$ color vs. rotation period $P_1$ of the \numonefilter stars with at least one detected period (gray dots). The filled circles represent data from \citet{Baliunas1996} (olive), \citet{Kiraga2007} (purple), \citet{Irwin2011} (red), and \citet{Amy2013} (blue). Towards cooler stars we found an increase in rotation periods with a steep rise around $B-V \approx 0.6$ supplying evidence of rotational braking. The black lines represent a color-period relation found by \citet{Barnes2007} for different isochrones. The four stars mark the position of the Sun (blue), KIC 1995351 (cyan), KIC 1869783 (purple), and KIC 9580312 (red) for comparison.}
  \label{all}
\end{figure*}

\begin{figure} 
  \resizebox{\hsize}{!}{\includegraphics{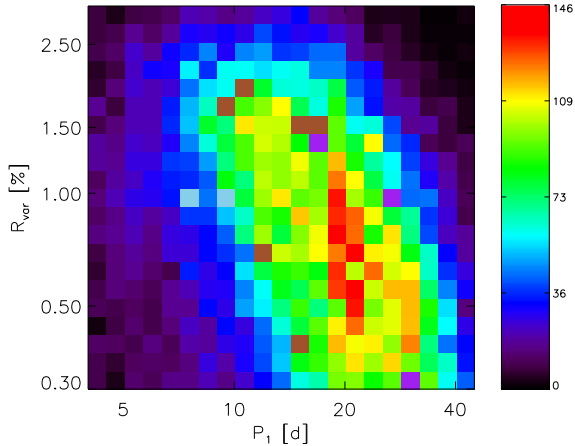}}
  \caption{Density plot of rotation periods $P_1$ vs. $R_{var}$ with bright region representing a high density, whereas dark regions express low density. The annotation of the color bar contains the total number of stars in each bin. The distribution shows that the range increases towards shorter periods. Since fast rotators are expected to exhibit an enhanced level of activity, the range could probably be used as proxy for stellar activity.}
  \label{densrange}
\end{figure}

\begin{figure}
  \resizebox{\hsize}{!}{\includegraphics{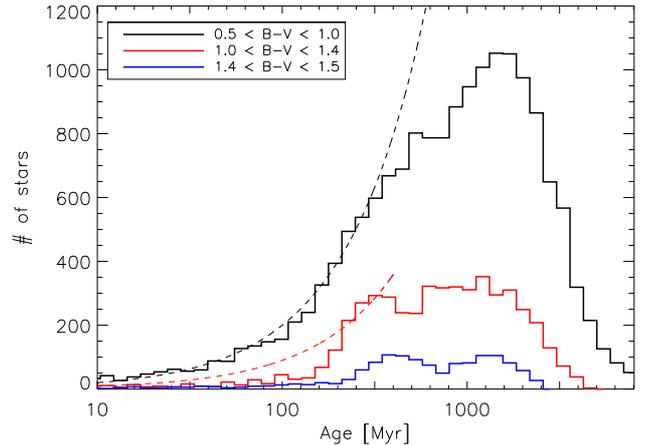}}
  \caption{Distribution of ages inferred from rotation periods $P_1$ for different color bins. For $B-V > 1.0$ (red and blue) we found a bimodal distribution in agreement with \citet{Amy2013}. The dashed black and red lines show uniform age distributions (on a logarithmic scale).}
  \label{ages}
\end{figure}

It is well-known that stellar rotation rates correlate with spectral type. Stars around spectral type F and earlier are known to be fast rotators. The convection zone grows towards cooler stars, and a dynamo mechanism generates magnetic fields. Ionized material follows the magnetic field lines (stellar wind) and carries away angular momentum resulting in a spin-down of the star \citep{Barnes2003, Reiners2012}. This process is known as rotational braking, for which we found evidence in Fig. \ref{all}.
We plotted our most significant period $P_1$ against $B-V$ for \numonefilter stars with at least one detected period (gray dots). Fig. \ref{all} is a composition of previous rotation studies and our results. Filled circles represent data from \citet{Baliunas1996} (olive), \citet{Kiraga2007} (purple), and \citet{Irwin2011} (red). Recently, rotation periods for the Kepler M dwarfs sub-sample were published by \citet{Amy2013} (blue circles). Gray dots represent our measurements. For some of the other measurements no $B-V$ information was available. We transformed stellar masses into effective temperature using 1\,Gyr isochrone models from \citet{Baraffe1998}. The temperatures have been transformed into $B-V$ using the relation from \citet{Reed1998}. For the Kepler stars we used the relation between $g-r$ and $B-V$ from \citet{Jester2005}. The periods from \citet{Baliunas1996} form an upper envelope to our results. The results for the Kepler M dwarfs (blue circles) show good agreement with our results (see also Fig. \ref{comp}), although \citet{Amy2013} used an auto-correlation method which is a completely different mathematical tool. \\
Fig. \ref{all} is consistent with the picture of rotational braking. A steep rise in rotation periods appears around $B-V\approx0.6$. In this region the convection zone starts to form, and grows deeper in cooler stars. Thus, magnetic braking becomes stronger leading to a spin-down of the stellar rotation rate. \citet{Barnes2007} empirically found a relation between $B-V$, age $t$, and rotation period
\begin{equation}\label{barnes}
 P(B-V, t) = 0.7725\,(B-V-0.4)^{0.601} \, t^{0.5189}.
\end{equation}
The age-dependence is similar to the classical Skumanich law $P\propto\sqrt{t}$ (see also \citet{Reiners2012}). The black curves in Fig. \ref{all} represent isochrones with ages of 100, 600, 2000, and 4500\,Myr. The distribution of periods in our sample follows a similar behavior as the isochrones indicating that stars with different color follow similar age distributions. Stars with rotation periods around 5 days are probably very young stars, which had no time to spin-down. In Fig. \ref{densrange} we show rotation period $P_1$ vs. range $R_{var}$ in a density plot. The bright region in the middle represents a high density, whereas dark regions express low density. Young stars are not only expected to rotate fast but, also to be very active. We found that the range increases toward fast rotation, supporting this relation between our measured rotation period and variability range. Hence, the variability range is a useful age and activity indicator.
\\
We estimated the ages from rotational period for the Kepler stars using eq. (\ref{barnes}). The distribution of ages in the active Kepler stars is shown in Fig. \ref{ages}. The black histogram shows all stars with $0.5 < B-V < 1.0$, the red one covers $1.0 < B-V < 1.4$, and the blue one shows $1.4 < B-V < 1.5$. The steep decrease on the right-hand side of all three distributions can be understood by the missing long-period stars caused by our upper limit of 45 days, and by our selection of active stars only. The left-hand side of the age-distribution is expected to be rather complete, because the lower limit of detectable periods of 0.5 days is not relevant even for very young active stars. The dashed black line shows the distribution of stars according to a uniform distribution of stellar ages (plotted on a logarithmic scale). The dashed curve is remarkably similar to the left-hand side of the black distribution up to an age of $\approx 300$\,Myr. \\
\citet{Amy2013} found evidence for a bimodal period distribution in the Kepler M dwarf sample $(1.21 < B-V < 1.62)$. The bimodality is explained by the existence of two distinct stellar populations. The gap at $P_1 \approx 30$ days also appears in our data around $1.4 < B-V < 1.5$, corresponding to an age of roughly 800--900\,Myr. We found a similar feature in hotter stars $(1.0 < B-V < 1.4)$ at $P_1 \approx 20$\,d. This feature, however, corresponds to a significantly younger age of 600--800\,Myr, and is unlikely to be caused by the same age distribution as the gap in cooler stars. Whether the two period gaps are caused by a selection bias affecting our period sample, or by a predominance of certain ages in the distribution of stellar ages (potentially introduced by stellar clusters in the Kepler field, see e.g. \citet{Meibom2011}) needs to be further tested. Furthermore, in comparison to a constant star formation history (dashed red line in Fig. \ref{ages}), the sample with $1.0 < B-V < 1.4$ (red histogram) lacks stars younger than $\approx 200$\,Myr. The dearth of young objects as well as the gaps in the age distributions cannot be easily explained and need further investigation. We leave this discussion for a subsequent paper.

\subsection{Relative differential rotation $\alpha$}\label{DR}
\begin{figure} 
  \resizebox{\hsize}{!}{\includegraphics{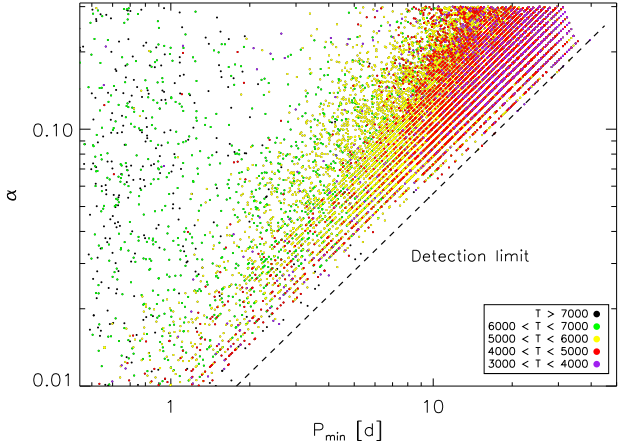}}
  \caption{Rotation period $P_{min}$ vs. $\alpha$ for all stars with two detected periods. The colors represent different temperature bins, the dashed black line marks the detection limit. The relative shear $\alpha$ grows towards longer rotation periods. More than 75\% of all stars with detected period $P_1$ exhibit a second period $P_2$, and lie above the detection limit. This trend cannot be broken by the \oneper stars with only one detected period, because most of the ``missing'' stars lie above the upper limit $\alpha_{max}=0.3$.}
  \label{detlim}
\end{figure}

\begin{figure}
  \resizebox{\hsize}{!}{\includegraphics{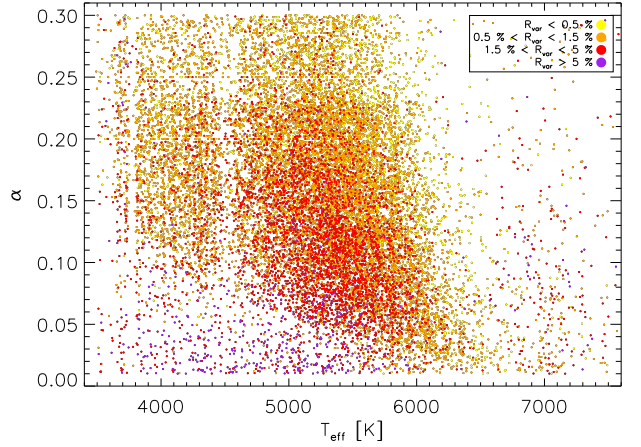}}
  \caption{Effective temperature vs. relative shear $\alpha$. We found that $\alpha$ slightly increases towards cooler stars. The colors indicate different variability ranges. A distinct correlation between $\alpha$ and the range is only visible for the stars with very high ranges $R_{var} > 5\%$. These stars group at small $\alpha$ values over a large temperature range. Low $\alpha$ represents short periods, probably indicating that their young age.}
  \label{teff_alpha_range}
\end{figure}

In Fig. \ref{detlim} we plot the relative DR $\alpha$ against the minimum rotation period $P_{min}$.
We found that $\alpha$ increases with rotation period. The black dashed line marks the detection limit (sec. \ref{detlimsec}) of our method. If we sort both periods that $P_1=P_{min}$ and $P_2=P_{max}$ then the absence of data points below this line can be understood by considering the relation
 \[|P_1-P_2|=\frac{|f_1-f_2|}{f_1 f_2}\geq 10 f_{low} P_1 P_2 \]
 \[\Rightarrow \quad \alpha=\frac{|P_1-P_2|}{P_{max}}\geq 10 f_{low} P_{min}\]
Thus, all data points lie above the black line, it being the lower limit for $\alpha$. It is worth noting that more than 75\% of all stars with detected period $P_1$ lie above the detection limit, because \numtwofilter of \numonefilter stars exhibit a second period. In other words, only the \oneper stars where only one period was detected can either lie below the detection limit, or above (below) our $\alpha_{max} (\alpha_{min})$ values, respectively. In the appendix we discuss different $\alpha_{max}$ values, and show that periods are found for $\alpha_{max}>0.3$ (see Fig. \ref{P_alpha}), which were discarded in some cases (see Fig. \ref{kepler_lcs_perdgms1}). Lowering $\alpha_{min}$ did not yield much different periods, so we conclude that the general trend of increasing $\alpha$ with rotation period was not biased by our detection limit. \\
The colors in Fig. \ref{detlim} represent different temperature bins. Towards cooler stars $(T_{\text{eff}} < 5000$\,K, red and purple dots) the rotation period increases, confirming the result from Fig. \ref{all}. Hot stars above 6000\,K (green and black dots) mostly populate the short periods covering the whole $\alpha$ range. This region is probably biased by pulsators, which was considered in sec. \ref{FP}. \\
In Fig. \ref{teff_alpha_range} we show temperature vs. $\alpha$, correlating our results with the variability range. The colors indicate different ranges growing from 0.3\% (yellow) to high ranges with amplitudes above 5\% (purple). A shallow trend towards higher $\alpha$ with decreasing temperature is visible (see also Fig. \ref{teff_alpha}). A correlation with range can only be found for stars with very high ranges (purple dots). These stars mostly cover the region with $\alpha\lesssim0.05$ over a large temperature range. If we think of the variability range as activity indicator this might confirm the hypothesis that these stars are very young, because low DR $\alpha$ indicates short periods.

\subsection{Absolute horizontal shear $\dom$}\label{DR1}
We define the absolute horizontal shear as 
\[ \dom:=2\pi|f_1-f_2| \]
\begin{figure}
  \resizebox{\hsize}{!}{\includegraphics{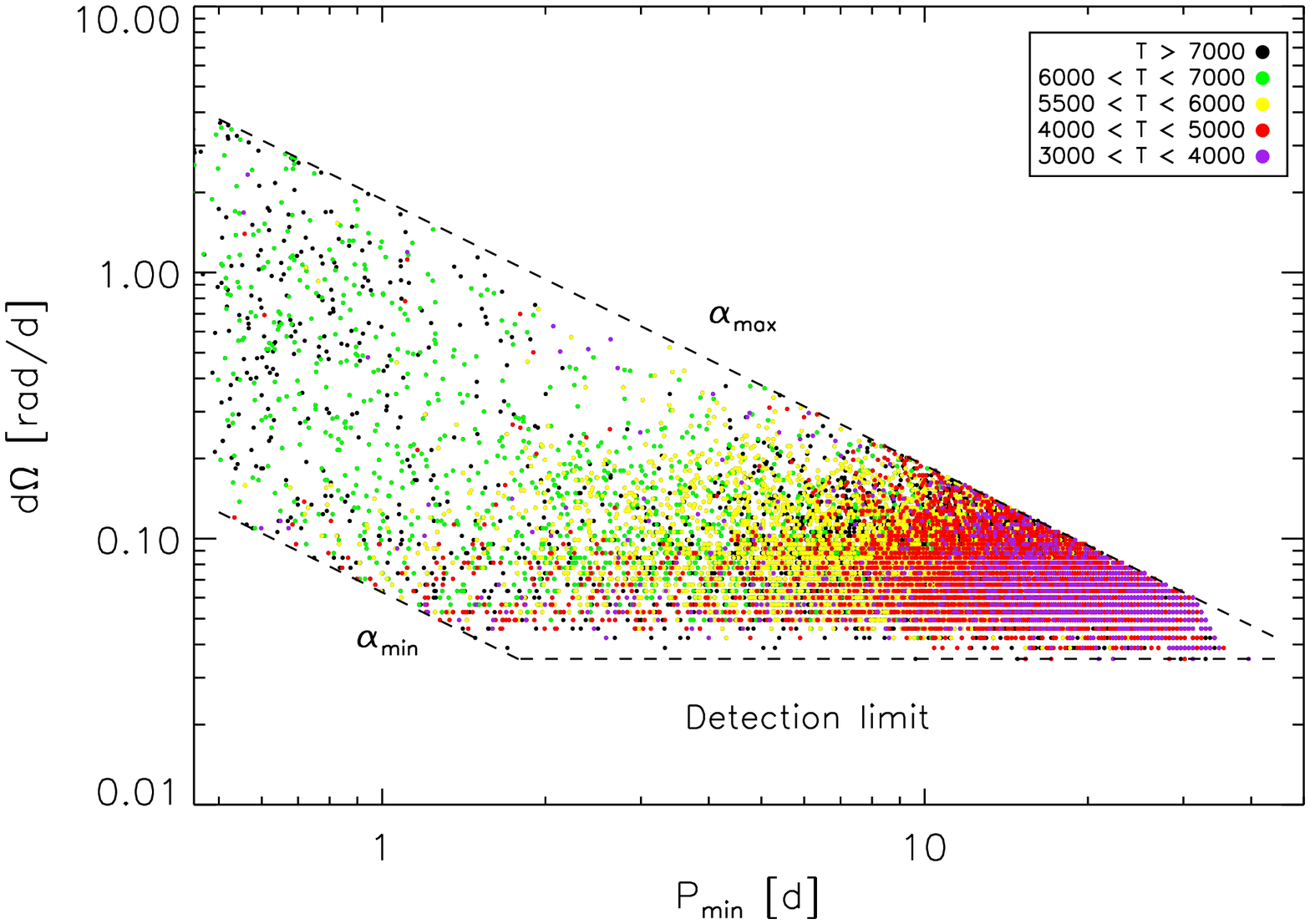}}
  \caption{Rotation period $P_{min}$ vs. absolute shear $\dom$. The colors are the same as in Fig. \ref{detlim}, the detection limit is marked by the horizontal dashed line, and the diagonal dashed lines mark the upper and lower limit $\alpha_{max}$ and $\alpha_{min}$, respectively. We found that $\dom$ is independent of $P_{min}$ over a large period range, although the upper and lower limit suggest an increase of $\dom$ towards short periods, showing large scatter on one order of magnitude.}
  \label{P1_dOmega}
\end{figure}
In Fig. \ref{P1_dOmega} we plot $\dom$ against the rotation period $P_{min}$. The colors are the same as in Fig. \ref{detlim}, the detection limit is marked by the horizontal dashed line, and the diagonal dashed lines mark the upper and lower limit $\alpha_{max}$ and $\alpha_{min}$, respectively. For periods longer than two days, $\dom$ shows weak dependence on rotation period. Below two days, $\dom$ shows large scatter on the order of one magnitude. The upper and lower limit suggest that $\dom$ increases toward short periods, which is not necessarily the case (compare Fig. \ref{Pmin_dOmega_dens}).
\begin{figure}
  \resizebox{\hsize}{!}{\includegraphics{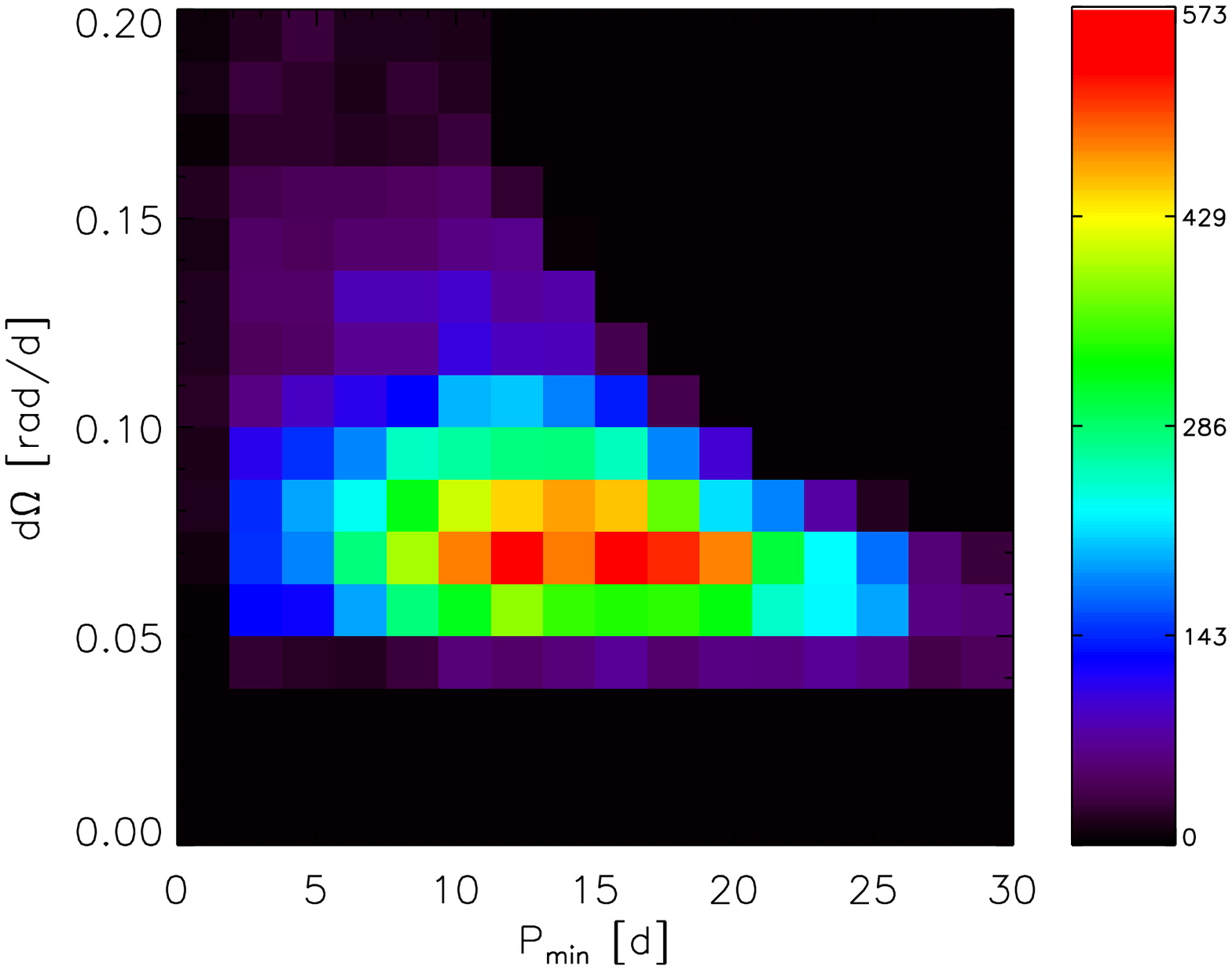}}
  \caption{Density plot in the $P_{\text{min}}-\dom$ plane. For $\dom < 0.10$\,rad\,d$^{-1}$ the total shear does not depend rotation period. For $\dom>0.10$\,rad\,d$^{-1}$ we found large scatter of $\dom$ from fast to moderate rotators.}
  \label{Pmin_dOmega_dens}
\end{figure}
Previous observations \citep{Barnes2005} and theoretical approaches \citep{kueker2011} also showed weak dependence of $\dom$ on rotation period. To point out this result in our measurements, we show a density plot in the $P_{\text{min}}-\dom$ plane in Fig. \ref{Pmin_dOmega_dens}. In the range $0.035\,\text{rad\,d}^{-1}\lesssim\dom\lesssim 0.10$\,rad\,d$^{-1}$ the absolute shear does not depend on rotation period. Above $0.10\,\text{rad\,d}^{-1}$ the shear shows large scatter for periods $0.5\,\text{d} \lesssim P_{\text{min}} \lesssim 15\,\text{d}$. Again, the upper limit $\alpha_{max}$ (compare Fig. \ref{P1_dOmega}) suggests a trend toward fast rotators. \\
Fig. \ref{Pmin_dOmega_dens} is similar to Fig. 3 in \citet{kueker2011} showing the dependence of $\dom$ on rotation period for different stellar mass models. These authors found weak dependence of $\dom$ on rotation period for all masses. Their model curves would fit our observations in Fig. \ref{Pmin_dOmega_dens}, although the $0.5\,\Msun$ and $0.3\,\Msun$ models lie below our detection limit. \citet{kueker2011} did not consider high DR $\dom > 0.1$\,rad\,d$^{-1}$, but their 1.1\,$\Msun$ model exhibits the largest value of $\dom$, and deviates the most from the almost constant shape of the other models, which could be a hint for a different behavior in this regime. \\
The temperature dependence of $\dom$ is shown in Fig. \ref{Teff_dOmega}, with our measurements plotted as gray dots. The olive diamonds and error bars are measurements, which were analyzed in \citet{Barnes2005}. The orange data and error bars are measurements from \citet{Ammler2012}. The olive dashed line shows the power-law fit from \citet{Cameron2007} combining data from \citet{Barnes2005} with new measurements. The red dash-dotted line and the light blue dashed line are theoretical predictions from \citet{kueker2011}. The black dashed line marks our detection limit. The four stars mark the position of the Sun (blue), KIC 1995351 (cyan), KIC 1869783 (purple), and KIC 9580312 (red) for comparison. The vertical red bar indicates stars hotter than 7000\,K. In this region the derived periods are highly biased by pulsators (see sec. \ref{FP}). \\
\begin{figure*}
  \centering
  \includegraphics[width=17cm]{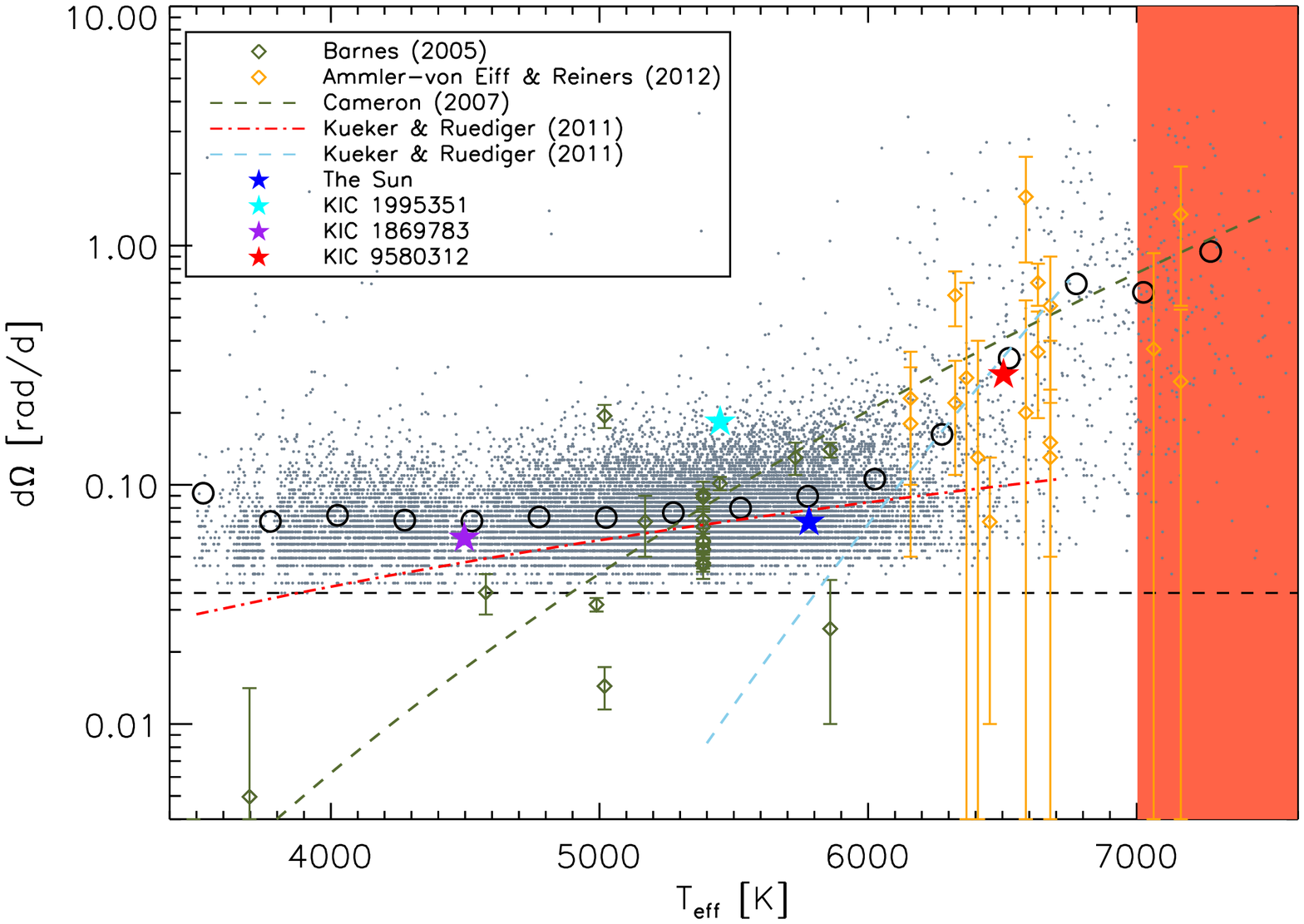}
  \caption{Effective temperature vs. horizontal shear $\dom$ summarizing different measurements: The olive diamonds and error bars were taken from \citet{Barnes2005}, the olive dashed curve was taken from \citet{Cameron2007}. Orange diamonds and error bars show measurements from \citet{Ammler2012}. Our measurements were shown as gray dots. The red dash-dotted line and the light blue dashed line show theoretical predictions from \citet{kueker2011}. The black dashed line marks our detection limit. The four stars mark the positions of the Sun (blue), KIC 1995351 (cyan), KIC 1869783 (purple), and KIC 9580312 (red)  for comparison. The vertical red bar indicates stars hotter than 7000\,K, with probably large contamination of pulsators. The open black circles represent weighted means of our measurements for different temperature bins. From 3500--6000\,K $\dom$ shows only weak dependence on temperature. Above 6000\,K the shear strongly increases as reported by \citet{Barnes2005,Cameron2007}. The different behavior of $\dom$ in these two temperature regions was supported by theoretical predictions from \citet{kueker2011} (red and light blue lines).}
  \label{Teff_dOmega}
\end{figure*}
Our results reveal two distinct temperature regions showing different behavior of $\dom$. For temperatures ranging from 3500--6000\,K $\dom$ shows only weak dependence on temperature. Above 6000\,K $\dom$ steeply increases toward hotter star, but with large scatter supporting no conclusion about the functional form of a fit. \citet{Barnes2005} found a strong temperature dependence of $\dom$ supplying the power-law $\dom\propto T_{\text{eff}}^{8.92}$. This was confirmed by \citet{Cameron2007} supplying the power-law fit $\dom=0.053(T_{\text{eff}}/5130)^{8.6}$. \\
Our results show good agreement with theoretical curves provided by \citet{kueker2011}. These authors found that the temperature dependence of $\dom$ cannot be represented by a single power-law fit, but requires two fits for different temperature regions (red and light blue curve, compare Fig. 2 in \citet{kueker2011}), which was clearly confirmed by our measurements. For temperatures of 3500--6000\,K \citet{kueker2011} predict a shallow increase of $\dom$ (red dash-dotted line). Since this behavior is not evident in Fig. \ref{Teff_dOmega}, we calculated histograms of $\dom$ for temperature bins of 250\,K between 3400\,K and 7400\,K. For each distribution, i.e. for each temperature bin, the weighted mean $\left\langle \dom\right\rangle$ was calculated, and is shown as open black circle in Fig. \ref{Teff_dOmega}. The mean values slightly increase for temperatures of 3500--6000\,K. Above 6000\,K $\left\langle \dom\right\rangle$ shows very good agreement with the light blue dashed curve, as predicted by \citet{kueker2011}. For even hotter stars $(T_{\text{eff}} > 7000$\,K) the derived periods are most probably highly contaminated by pulsators (see sec. \ref{FP}). In this region the derived periods and $\dom$ values should be treated with caution, which was indicated by the vertical red bar. \\
Although our measurements suggest good agreement between theory and observations in both regions, the results should be treated with caution. For the \oneper stars, for which only one period was detected, we cannot tell whether these stars exhibit a small horizontal shear below below our detection limit, or if a second period was not detected for other reasons. A fraction of these \oneper stars can change the trends we found, so we cannot draw strong conclusions about the behavior below the lower limit $\dom \approx 0.035$\,rad\,d$^{-1}$.

\subsection{False Positives}\label{FP}
In this section we statistically estimated the number of periods in our final sample, which survived all filters but are probably not due to rotation. Other sources of periodic stellar variability are e.g. binarity, pulsations, or instrumental effects. We call these detections false positives (FPs). We defined three classes of rotational variability: the rapid rotators $(0.5\,\text{d} < P_1 < 10\,\text{d})$, the moderate rotators $(10\,\text{d} < P_1 < 20\,\text{d})$, and the slow rotators ($P_1 > 20$\,d). Example light curves representative for the first and last group are shown in Fig. \ref{kepler_lcs_perdgms} and \ref{kepler_lcs_perdgms1}. Fig. \ref{Fstar} shows an example of an F-star. These hot stars were investigated separately at the end of this section. Since the sample is too large to inspect each light curve individually, we randomly selected 100 stars of each rotation class and checked their light curves by eye. This will only give a rough error estimate, because only a small number of stars was inspected, and the method remains very subjective. \\
Orbital periods of eclipsing binaries or transiting planets should be relatively rare in the final sample. The eclipsing binary list\footnotemark[3] was cross-matched with our sample, and stars with coinciding KIC numbers were discarded. For stars hosting planets or planetary candidates, the returned periods were most dominantly due to stellar activity. The transits cover very few data points compared to rotational modulations, hence the periodogram is not very sensitive to these periods. We found no period associated to eclipsing binaries or planetary transits in this test. \\
The rapid rotators class exhibits six stars showing irregular variations, which could be a hint for stellar pulsations (see below), six alias periods, and two periods without any reference to the light curve. \\ 
For the moderate rotators class alias periods were the biggest error source. In six cases the detected period was most probably the half of the true rotation period, and in one case the determined period could not be confirmed by visual inspection of the light curve. \\ 
The slow rotators class with periods greater than 20 days mostly suffers from instrumental effects. After each third of a quarter ($\approx 30$ days), data is down-linked to Earth cutting the quarter into three segments. Each raw data segment shows individual trends, which were corrected by the PDC-MAP pipeline. Thus, on time scales longer than 30 days, the accuracy of the derived periods strongly depends on the data reduction pipeline. Furthermore, we expect that star spots may evolve on the time scales, as seen on the Sun for stars with solar rotation periods, which distorts the light curve shape and makes it more difficult to detect stable periods. In this class only one alias period was detected, but 12 stars that did not show clear counterparts to the rotation period in the light curve (like regular dips or double dips as seen for fast rotating stars). \\ 
In total 300 stars were inspected. Summarizing results for the above rotation classes, we found that about 12\% of all derived periods should probably be attributed to sources of periodic variability other than rotation. \\
As evident from Fig. \ref{detlim}, stars hotter than 6000\,K mostly exhibit periods less than 1--2 days. This regime is highly biased by pulsations \citep{Debosscher2011}, because F-type and hotter stars exhibit very thin convection zones. $\delta$ Scuti stars show pulsations on timescales of less than half a day. Thus, we set a lower limit of 0.5 days to our periods in sec. \ref{selection}. But there do exist hot A-type stars with $\gamma$ Doradus pulsations with periods of 0.5--4\,d \citep{Balona2011_Astars,Balona2011_gammaDor,Balona2013}. Moreover, these stars show beat-shaped light curves similar to spot-induced variability. To check whether the periods found are due to rotation or pulsation, we defined three groups of hot stars, following the classification of Table 1 in \citet{Balona2011_Astars}: F9-F5 stars with $6000\,\text{K} < T_{\text{eff}} < 6500$\,K, F5-F1 stars with $6500\,\text{K} < T_{\text{eff}} < 7000$\,K, and hotter stars with $T_{\text{eff}} > 7000$\,K. Again, for each group we randomly selected 100 stars, and checked their variability by eye. If a traveling wave was found in the light curve, which was visible over several periods, we interpreted this behavior as migrating spots (compare Fig. \ref{Fstar}) indicating DR. \\
For the F9-F5 stars the derived periods are mostly rotation-induced. Only six stars were mis-classified, two most probable pulsating stars, three stars showing irregular variability, and one eclipsing binary were detected. In the second group (F5-F1 stars) beat-shaped light curves are found quite frequently. In thirty-three stars no traveling wave was found, although some of them exhibit a beat-shaped light curve. Thus, the variability probably arises from pulsations. \\
The situation becomes even worse for the hottest stars. Thirty-seven stars without moving dips were found, five periods without clear reference in the light curve, and three alias periods. This group shows the biggest contamination by $\gamma$ Dor or extreme $\delta$ Scuti pulsators. Although it is challenging to distinguish between rotation and pulsation from the light curve alone, short periods of hot stars should be treated with caution.

\section{Discussion}\label{discussion}
\subsection{Comparison to other observations}
The rotation periods of the active Kepler stars are consistent with previous rotation measurements (Fig. \ref{all}), supporting the picture of stars losing angular momentum due to stellar winds, which was deduced from a long history of observations. 
\begin{figure}
  \resizebox{\hsize}{!}{\includegraphics{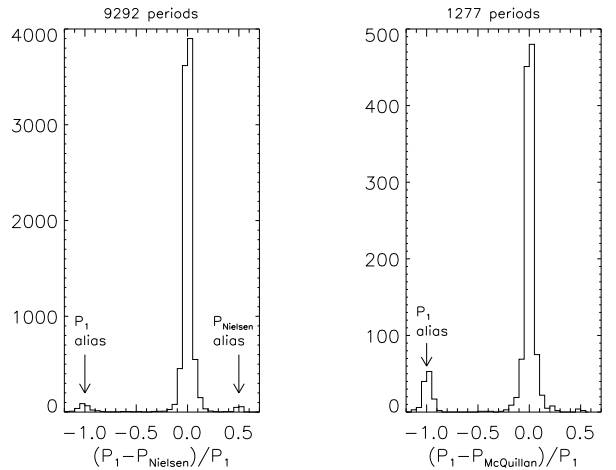}}
  \caption{Comparison of periods from \citet{Nielsen2013} (left panel) and \citet{Amy2013} (right panel) with the periods $P_1$ from Fig. \ref{all}. We found good agreement with both samples, although the auto-correlation method from \citet{Amy2013} was less prone to alias periods than our method, as obvious by the missing peak around $0.5$ in the right panel.}
  \label{comp}
\end{figure}
Recently, \citet{Nielsen2013} and \citet{Amy2013} presented rotation periods for a large sample of Kepler stars. To create confidence in our results and the method we used, we compared our periods $P_1$ to the periods derived by these authors. The left panel in Fig. \ref{all} shows the normalized period difference $(P_1-P_{\text{Nielsen}})/P_1$. In total 9292 periods were compared. The huge peak around zero shows that our periods are consistent with the periods derived in \citet{Nielsen2013}, which were derived from analyzing multiple Kepler quarters. The small peak at $-1$ indicates the cases where we detected an alias period $P_1=P_{\text{Nielsen}}/2$. The small peak around $0.5$ shows the opposite effect when \citet{Nielsen2013} detected an alias according $P_{\text{Nielsen}}=P_1/2$. Both peaks are of the same size as expected, because the Lomb-Scargle periodogram was used in both studies. \\
The right panel in Fig. \ref{all} shows the normalized period difference $(P_1-P_{\text{McQuillan}})/P_1$ comparing 1277 periods in total. Again, we found good agreement between both samples. The peak at $-1$ indicating the cases where we detected an alias period $P_1=P_{\text{McQuillan}}/2$ is rather large compared to the peak around zero. This was also expected since M-dwarfs exhibit long rotation periods bearing the risk of aliasing, especially when using an automated method. Around $0.5$ there is no peak visible, clearly demonstrating that the auto-correlation method used in \citet{Amy2013} was less prone to alias periods than our method. \\
Our results on DR are quite different than what was observed before. In contrast to previous observations, e.g. \citep{Barnes2005,Cameron2007}, we found that $\dom$ weakly depends on temperature for the cool stars (3000--6000\,K). Above 6000\,K $\dom$ increases with temperature but the stars show no systematic trend, but seem to be randomly distributed in this temperature regime. Using Doppler imaging (DI) \citet{Barnes2005} found that the horizontal shear $\dom$ strongly depends on effective temperature $\dom\approx T_{\text{eff}}^{8.92}$. Five stars of their sample lie below our detection limit (compare Fig. \ref{Teff_dOmega}). \citet{Cameron2007} combined results from DI and the Fourier transform method yielding the equation $\dom=0.053\,(T_{\text{eff}}/5130)^{8.6}$. The two groups we found (above and below 6000\,K) were also consistent with recent theoretical studies (compare sec. \ref{theory}). \\
The relation between rotation period and DR has been studied by several authors. \citet{Hall1991} found that the relative horizontal shear $\alpha$ increases towards longer rotation periods. \citet{Donahue1996} confirmed this trend finding $\Delta P \approx \left\langle P \right\rangle^{1.3\pm0.1}$, independent of the stellar mass. Using the Fourier transform method \citet{Reiners2003} also found that $\alpha$ increases with rotation period for F-G stars. This result was confirmed by \citet{Ammler2012} compiling previous results and new measurements for A-F stars. \citet{Barnes2005} found that $\dom$ only weakly correlates with rotation rate according $\dom\approx\Omega^{0.15}$. \\
Observations of DR cover a wide range of $\alpha$ values. DI is particularly sensitive to small DR limited by the spot lifetimes, $\alpha \lesssim 0.01$ (see e.g. measurements by \citet{Donati1997} for AB Dor), although there were measurements \citep{Donati2003} yielding $\alpha\approx 0.05$ for LQ Hya, and new measurements \citep{Marsden2011}, who found values between $0.005\lesssim\alpha\lesssim 0.14$. The Fourier transform method (e.g. \citet{Reiners2003}) is sensitive to $\alpha > 0.1$ and was used to determine surface shears as large as 50\% for some A-F stars. With our method we are able to detect DR up to $\alpha<0.5$. The errors for individual periods are usually small. Bad detections of $\alpha$ result from the spot distribution on the stellar surface, which is a general problem for DR detections from photometric data. The measured shear $\alpha$ will always be lower than the total equator-to-pole shear. The accuracy of our method was discussed in \citet{Reinhold2013}. \\
In the following we compared our results with previous rotation measurements of individual Kepler stars (compare table \ref{kic_table_v2}). For three Kepler stars (KIC 8429280, KIC 7985370, KIC 7765135) DR has been measured \citep{Frasca2011,Froehlich2012} fitting an analytical spot model to the data. Their findings showed good agreement to our results and have been compared in Paper I. \\
Several Kepler stars were measured by Savanov using a light curve inversion technique that constructs a map of surface temperature. \citet{Savanov2011_KOI} considered the planet-candidate host stars KOI 877 and KOI 896 (KIC 7287995 and KIC 7825899, respectively) finding rotation period of 13.4 and 25.2 days, respectively, stating that both stars exhibit active longitudes separated by about 180$\degree$. For KIC 7287995 we found $P_1=13.5$\,d and $P_2=10.4$\,d, and for KIC 7825899 $P_1=12.4$\,d. In the latter case the detected period was an alias of the true rotation period due to the active longitudes. \\
The K dwarf KIC 8429280 was studied in \citet{Savanov2011_Kdwarf}, who found brightness variations with periods of 1.16 and 1.21 days, consistent with the results from \citet{Frasca2011}. In this case we found $P_1= 1.16$\,d and $P_2= 1.21$\,d. \\
The fully convective M dwarf GJ 1243 (KIC 9726699) was studied in \citet{Savanov2011_Mdwarf} yielding a rotation period of 0.593 days. Our algorithm detected a period of 118.7 days due to improper data reduction in Q3. This period was filtered out by the upper limit of 45 days. The second strongest period we found was 0.59 days, in good agreement with \citet{Savanov2011_Mdwarf}. This period does not lie within 30\% of the first one, and hence was not chosen as $P_2$. \\
\citet{Savanov2012} also studied the fully convective, low-mass M dwarf LHS 6351 (KIC 2164791). These authors detected a rotation period of 3.36\,d, and evidence for DR in terms of $\dom=0.006-0.014$\,rad\,d$^{-1}$ from the evolution of surface temperature inhomogeneities, which lies below our detection limit. We found $P_1=3.35$\,d and $P_2=3.27$\,d yielding $\dom=0.046$\,rad\,d$^{-1}$. Visual inspection of the light curve in Q3 supports the larger value of $\dom$, because a second active region appears after some periods, which was not observed in Q1-Q2 data. Unfortunately, this star misses in our final list because it has no effective temperature or $\log$g values from the KIC. \\
\citet{Bonomo2012} analyzed the active planet-hosting stars Kepler-17 (KIC 10619192). These authors detected a rotation period of 12.01\,d, and the existence of two active longitudes separated by approximately $180^\circ$ from each other. Furthermore, \citet{Bonomo2012} found evidence for solar-like DR,  but were not able to give precise estimates for the horizontal shear claiming that active regions on this star evolve on timescales similar to the rotation rate. We detected an alias period of 6.05\,d due to the two active longitudes. Again, we found no second period within 30\% of $P_1$. Checking our second and third strongest period, we found 10.85\,d and 12.28\,d, respectively, in good agreement with rapid spot evolution. \\
Recently, \citet{Roettenbacher2013} found a period of 3.47 days for the Kepler target KIC 5110407, and evidence for DR using light curve inversion. We found $P_1=3.61$ and $P_2=3.42$ days yielding $\alpha=0.053$, which was consistent with their differential rotation coefficient $k=0.053\pm0.014$ for an inclination of $i=45\degree$ in their model. \\
Several open clusters in the Kepler field were studied. \citet{Meibom2011} measured rotation periods for stars in the open cluster NGC 6811, supplying evidence for rotational braking (compare Fig. \ref{all}). Further measurements were done for the open clusters NGC 6866 \citep{Balona2013_NGC6866} and NGC 6819 \citep{Balona2013_NGC6819} showing rotational braking as well. The same author found several A-type stars showing signatures of spot-induced variability \citep{Balona2011_Astars, Balona2013}, e.g. a beating pattern in the light curve, although this behavior was not expected due to the purely radiative atmospheres. 
\begin{table}
  \caption{Comparison with previous rotation measurements for individual Kepler stars. References in the third column: (\one) \citet{Frasca2011}, (\two) \citet{Froehlich2012}, (\three) \citet{Savanov2011_KOI}, (\four) \citet{Savanov2011_Kdwarf}, (\five) \citet{Savanov2011_Mdwarf}, (\six) \citet{Savanov2012}, (\seven) \citet{Bonomo2012}, (\eight) \citet{Roettenbacher2013}.}
  \label{kic_table_v2}
  \begin{center}
   \begin{tabular}{ccccc}
\hline
KIC & Period(s) & Ref. & $P_1$ & $P_2$  \\
    &  [d]      &      & [d] & [d] \\
\hline
8429280 & 1.16--1.20 & \one & 1.16 & 1.21 \\
7985370 & 2.84--3.09 & \two & 2.84 & 3.09 \\
7765135 & 2.40--2.57 & \two & 2.55 & 2.40 \\
\hline
 7287995 & 13.4 & \three & 13.5 & 10.4 \\
 7825899 & 25.2 & \three & 12.4 & - \\
 8429280 & 1.16, 1.21 & \four & 1.16 & 1.21 \\
 9726699 & 0.59 & \five & - & - \\
 2164791 & 3.36 & \six & 3.35 & 3.27 \\
10619192 & 12.01 & \seven & 6.05 & - \\
 5110407 & 3.47 & \eight & 3.61 & 3.42 \\
\hline
\end{tabular}
  \end{center}
\end{table}

\subsection{Comparison to theory}\label{theory}
Solar DR has theoretically been studied for a long time. \citet{Kitchatinov1999} computed DR models for late-type (G2 and K5) stars. They found that the relative shear $\alpha$ increases with rotation period. They also showed that $\alpha$ increases towards cooler stars. The general trends are in agreement with our findings, although their range of $\alpha$ values lies above our values. \\
\citet{Kueker2005} computed models for an F8 star, and found weak dependence on rotation period, which was confirmed by later studies for F, G and K stars \citep{Kueker2005_lowMS}, showing that the dependence on temperature was much stronger. The latter result holds for different main sequence star models \citep{Kueker2007}. These authors found that above a temperature of 5800\,K the strong temperature dependence on $\dom$ claimed in \citet{Barnes2005} fits the model data reasonably well, whereas below 5800\,K the data lie far off the fit. Recent studies \citep{kueker2011} have shown that the temperature dependence of $\dom$ cannot be represented by one single power law over the whole temperature range from 3800--6700\,K. Fig. 2 in \citet{kueker2011} shows that $\dom$ only slightly increases with temperature for stars cooler than $\approx$ 5800\,K consistent with our measurements (Fig. \ref{Teff_dOmega}). For stars hotter than $\approx$ 6200\,K these authors found that $\dom$ shows an even stronger dependence on temperature than predicted in \citet{Barnes2005}. In the same temperature region, our measurements start to show a different behavior (compare Fig. \ref{Teff_dOmega}). \\
The weak dependence of $\dom$ on rotation period was confirmed by our measurements (compare Fig. \ref{Pmin_dOmega_dens}), as predicted by \citet{kueker2011} for different solar mass models, and other authors before. \citet{Hotta2011} modeled DR of rapidly rotating solar-type stars and found that DR approaches the Taylor-Proudman state, i.e. that $\dom/\Omega$ decreases with angular velocity, as long as the rotation rate is above the solar value. This model agrees well with our findings in Fig. \ref{detlim}. \\
Direct numerical simulations showed that models with low Rossby numbers (i.e. fast rotating stars) generate strong dipolar magnetic fields \citep{Schrinner2012}. These fields amplify Lorentz forces able to suppress Coriolis forces, and hence can effectively suppress DR \citep{Gastine2012}. This result from theoretical models agrees well with the trend we found that $\alpha$ decreases towards shorter rotation periods (Fig. \ref{detlim}). \\
The dipolarity of the magnetic field strongly depends on the length scale of convection. As the depth of the convection zone decreases the dipolarity breaks down rendering the rotation non-uniform \citep{Schrinner2012}. This effect could explain the strong increase of $\dom$ around $T_\text{eff} > 6000$\,K. \citet{Browning2011} provided an explanation for the shallow increase of $\alpha$ towards cooler stars. M dwarfs exhibit low luminosities, and therefore low convective velocities. Thus, they strongly depend on the rotation rate even at solar rotation rates. Magnetic fields cannot quench DR effectively so they exhibit strong solar-like DR.

\section{Summary}\label{summary}
We applied our method from Paper I to the active fraction of Kepler Q3 data to search for DR in high precision empirical data. We measured rotation periods of \numonefilter Kepler stars, providing evidence for DR in \numtwofilter stars. Our measurements for the rotation period were in good agreement with previous results. Moreover they were consistent with the theory of magnetic braking. \\
Our measurements provide a comprehensive database of stellar DR. For the first time, we could explore a well-defined parameter range in a statistically significant sample. We found that the relative shear $\alpha$ increases towards longer periods, and slightly increases towards cooler stars. The absolute shear $\dom$ showed weak dependence on rotation period over a large period range. In contrast to other observations $\dom$ showed a shallow increase for temperatures from 3500--6000\,K, and a steep increase above 6000\,K. Periods above 7000\,K should be treated with caution due to probably high contamination of pulsators. Both the dependence on rotation period and temperature were in good agreement with recent theoretical models. Furthermore, we cannot find any other reasonable explanation for the trends we found. \\
We interpreted the existence of a second period as DR. Although our method is not able to distinguish between DR and spot evolution, we were confident that most of our measurements reflect the stellar surface shear because they resemble previous measurements and recent theoretical models. This was the first time that DR was measured for such a large number of stars. For the future this analysis will be applied to more Kepler data to verify the rotation periods, DR, and to learn about spot lifetimes. Kepler is a natural source of answers to these questions.

\bibliography{biblothek_v3}
\bibliographystyle{aa}

\acknowledgements TR acknowledges support from the DFG Graduiertenkolleg 1351 \textit{Extrasolar Planets and their Host Stars}. AR acknowledges financial support from the DFG as a Heisenberg Professor under RE 1664/9-1. 

\begin{appendix}
\section{Differential Rotation beyond $\alpha=0.3$}\label{beyond}
\begin{figure*}
 \centering
 \includegraphics[width=17cm]{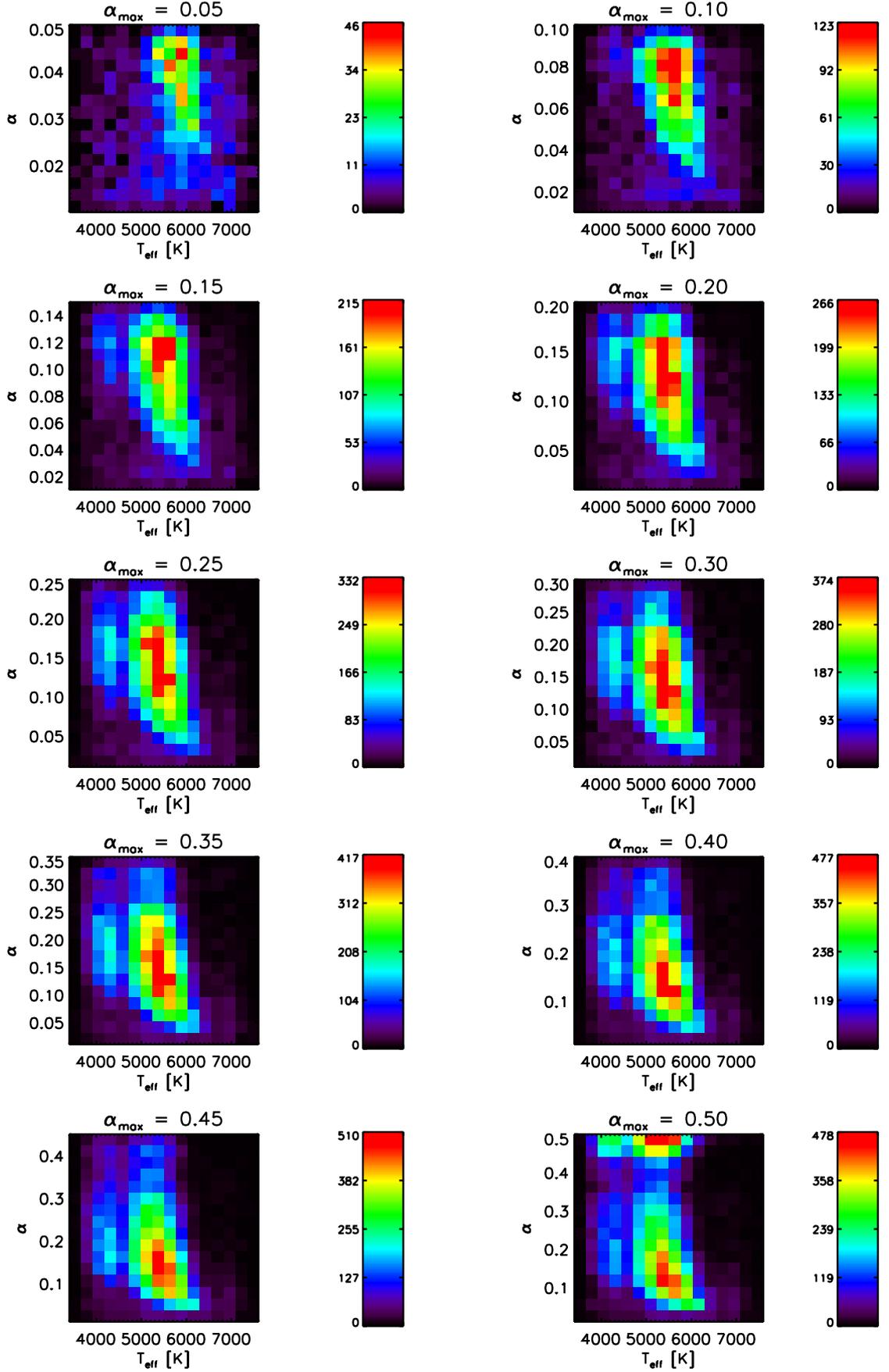}
 \caption{Density plot in the $T_{\text{eff}}-\alpha$ plane for different values of $\alpha_{max}$. For each $\alpha_{max}$ value we found that $\alpha$ slightly increases towards cooler stars. The plot in the lower right corner ($\alpha_{max}=0.5$) demonstrates that our method is limited to $\alpha_{max}<0.5$. Colors are the same as in Fig. \ref{densrange}.}
 \label{teff_alpha}
\end{figure*}

\begin{figure*}
 \centering
 \includegraphics[width=17cm]{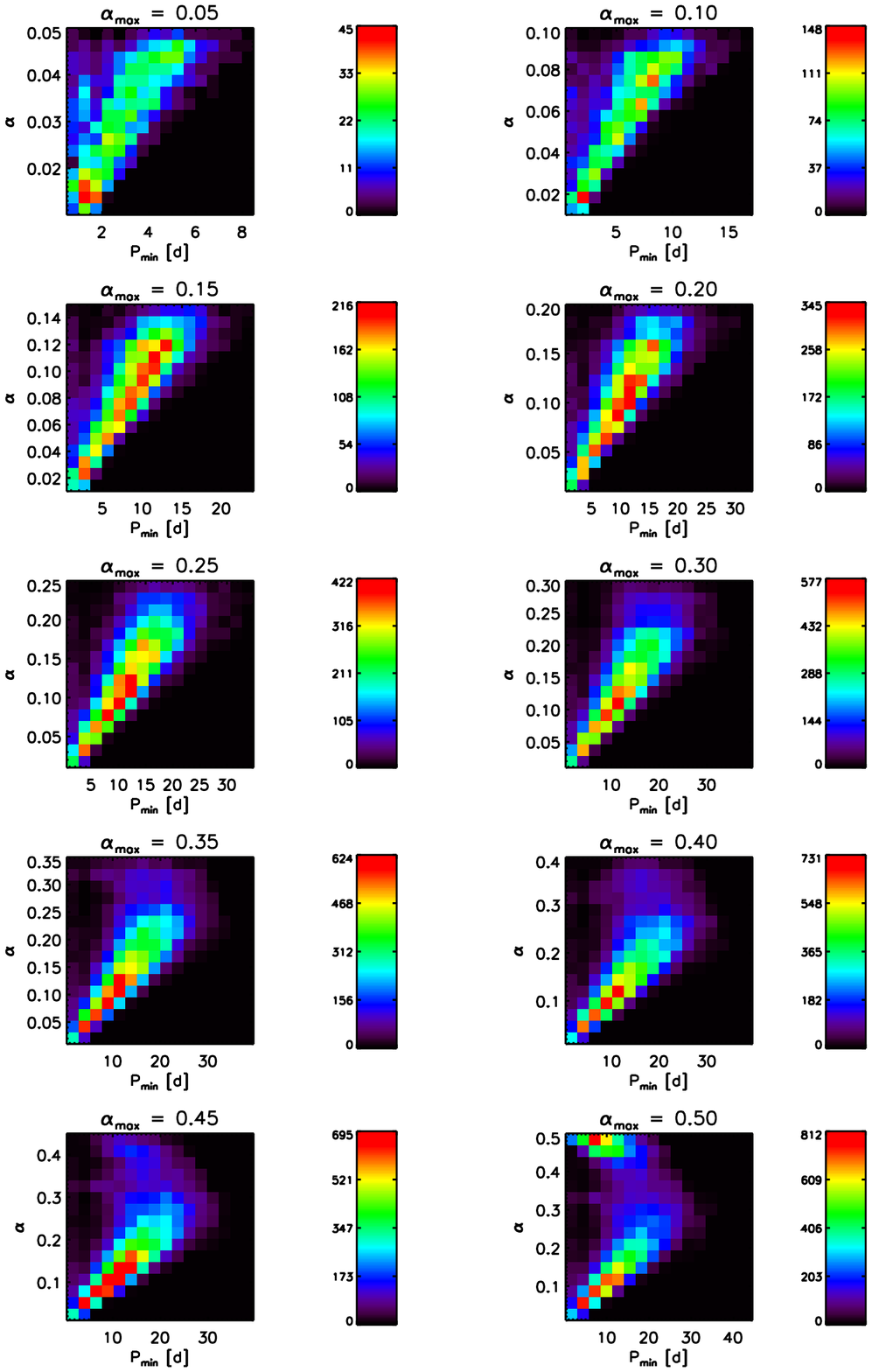}
 \caption{Density plot in the $P_{min}-\alpha$ plane for different values of $\alpha_{max}$. For each $\alpha_{max}$ value we found that $\alpha$ strongly increases with rotation period. The plot in the lower right corner ($\alpha_{max}=0.5$) demonstrates that our method is limited to $\alpha_{max}<0.5$. Colors are the same as in Fig. \ref{densrange}.}
 \label{P_alpha}
\end{figure*}

In eq. (\ref{limits}) an upper limit of $\alpha_{max}=0.30$ was used while searching for a second period. In this section we demonstrate how different upper limits $\alpha_{max}$ change the total number of detections and the overall behavior of $\alpha$ with temperature and rotation period. We use ten equidistant values $0.05\leq \alpha_{max} \leq 0.50$ (compare Table \ref{secper_table}).
\begin{table}
  \caption{Number of stars with second period found for different $\alpha_{max}$ values.}
  \label{secper_table}
  \begin{center}
    \begin{tabular}{cc}
      $\alpha_{max}$ [\%] & \# of two det. periods \\
      \hline
       5 & 1966 \\
      10 & 5437 \\
      15 & 9511 \\
      20 & 13355 \\
      25 & 16426 \\
      30 & 18619 \\
      35 & 20379 \\
      40 & 21559 \\
      45 & 22319 \\
      50 & 23206
    \end{tabular}
  \end{center}
\end{table}
With increasing $\alpha_{max}$ we found a larger total number of periods. The case $\alpha_{max}=0.5$ is shown to demonstrate the limits of our method, which is evident in the next two figures. For all stars with two detected periods we plotted their density in the $T_{\text{eff}}-\alpha$ plane in Fig. \ref{teff_alpha}. The colors are the same as in Fig. \ref{densrange} with bright regions representing a high density. For each $\alpha_{max}$ value we found that $\alpha$ slightly increases towards cooler stars. The case with the lowest upper limit ($\alpha_{max}=0.05$) looks a bit scattered but not very different than the general trend. The plot in the lower right corner ($\alpha_{max}=0.5$) demonstrates the limits of our method. Many stars ``jumped'' to the upper limit $\alpha=0.5$, because an alias period was chosen by the algorithm. The $T_{\text{eff}}$ values from the KIC are not very accurate and neither are the stellar radii. We found the same trend vs. $\alpha$ however, i.e. the $\alpha$ value increases towards smaller radii (although this may not be an independent constraint).
Our previous result that $\alpha$ increases towards longer rotation periods holds for all $\alpha_{max}$ values. In Fig. \ref{P_alpha} we showed the density in the $P_{min}-\alpha$ plane. We see that the density is localized in a sharp strip that smears out for limits greater than $\alpha_{max}=0.3$. These large $\alpha$ values belong to periods $P_2$ longer than 45 days, where instrumental effects play a dominant role. Again, the stars ``jumped'' to the upper limit in the $\alpha_{max}=0.5$ case.
\end{appendix}


\end{document}